\documentclass{aa}
\usepackage{bm}
\usepackage{breqn} % nessary to get functions in more than one line (it seems)

\usepackage{makecell}
\usepackage{graphics,txfonts}
\usepackage{natbib,twoopt}
\usepackage{subcaption}
\usepackage[breaklinks=true]{hyperref} %% to avoid \citeads line fills
\bibpunct{(}{)}{;}{a}{}{,}             %% natbib format for A&A and ApJ
\makeatletter
\newcommandtwoopt{\citeads}[3][][]{\href{http://adsabs.harvard.edu/abs/#3}%
	{\def\hyper@linkstart##1##2{}%
		\let\hyper@linkend\@empty\citealp[#1][#2]{#3}}}
\newcommandtwoopt{\citepads}[3][][]{\href{http://adsabs.harvard.edu/abs/#3}%
	{\def\hyper@linkstart##1##2{}%
		\let\hyper@linkend\@empty\citep[#1][#2]{#3}}}
\newcommandtwoopt{\citetads}[3][][]{\href{http://adsabs.harvard.edu/abs/#3}%
	{\def\hyper@linkstart##1##2{}%
		\let\hyper@linkend\@empty\citet[#1][#2]{#3}}}
\newcommandtwoopt{\citeyearads}[3][][]%
{\href{http://adsabs.harvard.edu/abs/#3}
	{\def\hyper@linkstart##1##2{}%
		\let\hyper@linkend\@empty\citeyear[#1][#2]{#3}}}
\makeatother

\def\code#1{\texttt{#1}}

\begin{document}

        \title{Merger identification through photometric bands, colours, and their errors.}

        \author{L. E. Suelves
                \inst{1}
                \and 
                W. J. Pearson\inst{1}
                \and 
                A. Pollo\inst{1}
        }

        \institute{National Centre for Nuclear Research, Pasteura 7, 02-093 Warszawa, Poland\\
                \email{luis.suelves@ncbj.gov.pl}}
        
        \date{}

        \abstract
%       Context.
        {}
%       Aims.
        {We present the application of a fully connected neural network (NN) for galaxy merger identification using exclusively photometric information. Our purpose is not only to test the method's efficiency, but also to understand what merger properties the NN can learn and what their physical interpretation is.}
%       Methods.
        {We created a class-balanced training dataset of 5\,860 galaxies split into mergers and non-mergers. The galaxy observations came from SDSS DR6 and were visually identified in Galaxy Zoo. The 2$\,$930 mergers were selected from known SDSS mergers and the respective non-mergers were the closest match in both redshift and $r$ magnitude. The NN architecture was built by testing a different number of layers with different sizes and variations of the dropout rate. We compared input spaces constructed using: the five SDSS filters: $u$, $g$, $r$, $i$, and $z$; combinations of bands, colours, and their errors; six magnitude types; and variations of input normalization.}
%       Results.
        {We find that the fibre magnitude errors contribute the most to the training accuracy. Studying the parameters from which they are calculated, we show that the input space built from the sky error background in the five SDSS bands alone leads to 92.64 $\pm$ 0.15 \% training accuracy. We also find that the input normalization, that is to say, how the data are presented to the NN, has a significant effect on the training performance.}
%       Conclusions.
        {We conclude that, from all the SDSS photometric information, the sky error background is the most sensitive to merging processes. This finding is supported by an analysis of its five-band feature space by means of data visualization. Moreover, studying the plane of the $g$ and $r$ sky error bands shows that a decision boundary line is enough to achieve an accuracy of 91.59\%. } 
    % 286

    \keywords{galaxies: evolution -- galaxies: interactions -- galaxies: photometry -- methods: data analysis -- methods: numerical}
    
    \titlerunning{Merger identification through photometric bands, colours, and their errors}
    \authorrunning{L. E. Suelves et al.}
    \maketitle
-

\section{Introduction}
\label{sect:Introduction}

The evolution and dynamics of the galactic population are key elements for modelling the Universe's history and structure. Within the cold dark matter model, the gravitational pull between galaxies inside collapsing dark matter halos can lead to mergers: hierarchical growth interactions in which one galaxy merges with another, combining their structure, stellar population, and interstellar dust \citep{1978MNRAS.183..341W, 2014ARA&A..52..291C, 2015ARA&A..53...51S}. Merging processes have been found to likely enhance the stellar formation in the daughter galaxy, as seen in studies of luminous infrared galaxies \citep{1985MNRAS.214...87J, 1996ARA&A..34..749S, 10.1111/j.1365-2966.2012.20425.x} or of active galactic nuclei activity \citep{2005Natur.433..604D}, and to significantly modify the shape of the original galaxies during the process \citep{1972ApJ...178..623T}. Mergers are currently thought to make up fewer than 10\% of the galaxies at low redshifts \citep{2017MNRAS.470.3507M, 2019ApJ...876..110D}, evolving up to 20\% in the redshift interval z $\epsilon$ [2,3] \citep{2014A&A...565A..10T}.

Identifying galaxy mergers is a challenging task: their appearance depends on the properties of the progenitors and the event can last $\sim$1 billion years \citep{10.1111/j.1365-2966.2008.14004.x}. Besides, the exact start and end of a merger are difficult to define. The most straightforward method is to do a visual inspection, \citep{2010MNRAS.401.1043D, 2007MNRAS.382.1415S}, which immediately becomes too time-consuming for the enormous size of upcoming galactic surveys such as Euclid \citep{2011arXiv1110.3193L} or LSST \citep{2019ApJ...873..111I}. Another common way of identifying mergers is the close-pair method. Close pairs are identified where two galaxies are close on the sky and have similar redshifts \citep{2000ApJ...530..660B, 1997ApJ...475...29P, 2002ApJ...565..208P, 2003MNRAS.346.1189L, 2004ApJ...617L...9L, 2005AJ....130.1516D, 10.1093/mnras/sty098, 2019ApJ...876..110D}. The close-pair method needs high-quality redshifts to achieve proper precision. Thus, it requires data from extensive spectrometric surveys. A third method is to make use of different morphological parameters that quantify the disturbances in the sources' shapes, such as the Gini coefficient \citep{2003ApJ...588..218A}, the second-order moment of the brightest 20\% of the light \citep[M$_{20}$$ $;][]{2004AJ....128..163L}, the asymmetry \citep{1996ApJS..107....1A, 2000ApJ...529..886C}, the concentration \citep{1985ApJS...59..115K, 1994ApJ...432...75A, 2000AJ....119.2645B, 2003ApJS..147....1C}, or the smoothness \citep{Takamiya_1999, 2003ApJS..147....1C}. 

The introduction of deep learning techniques has been the natural next step, given its recent development as the next generation of scientific modelling. They have the advantage of classifying large numbers of sources in a fraction of the time the methods above would demand. For galaxy morphology recognition, convolutional neural networks (CNNs) have been most commonly applied. CNNs carry out image recognition that can be performed on astronomical frames \citep{2015MNRAS.450.1441D}. They have been used for merger identification \citep{10.1093/mnras/sty3232, 2019A&A...626A..49P, 2019A&A...631A..51P, 2019ApJ...872...76N, 2019MNRAS.490.5390B, 2020ApJ...895..115F, 2020A&A...644A..87W}, including in combination with other methods such as transfer learning \citep{2018MNRAS.479..415A} or `ordinary' neural networks (NNs) using morphological parameters \citep{2022A&A...661A..52P}.

Regardless of the study subject, the main requirement of machine learning techniques is the availability of datasets from which the classification properties can be learnt. The best way to obtain such a dataset for merger identifications is to build it from a large correctly and visually classified galactic catalogue, such as the ones offered by Galaxy Zoo\footnote{\url{http://www.galaxyzoo.org/}} \citep{2008MNRAS.389.1179L}. Galaxy Zoo is a citizen science project in which amateur astronomers and volunteers employ part of their free time to classify the morphology of galaxy images provided by professional surveys. Over the years it has delivered a large catalogue of morphological classifications that have been used in a number of studies \citep[e.g.][]{ 2020MNRAS.491.1554W}\footnote{\url{https://www.zooniverse.org/about/publications}}.

This paper aims to understand if it is possible to apply a machine learning model for merger identification taking only the photometric galactic measurements into account, without relying on astronomical images or morphological parameters. Therefore, the goal of our paper is to give a first view of how an NN of this type could perform, to test its potential, and find the properties it can learn from the data.

The paper is arranged as follows. In Sect. 2 we define the labelled dataset employed for training the NN and its applicable features. Then we address the methods applied in Sect. 3, mainly explaining the NN's internal specifications, together with other techniques we considered for analysing the inptut data. Section 4 covers the main results of our experiment, from the selected NN's architecture to the discovery of the successful role the sky background errors play in the identification. Section 5 discusses these results, and finally we conclude and discuss future avenues of work in Sect. 6.

\section{Dataset}
\label{sect:Dataset_selection}
The dataset used for training an NN is required to contain a representative sample for each different class intended to be identified. In this project, we are discerning merging and non-merging galaxies. Our selection of galaxy mergers was taken from the catalogue created by \cite{2010MNRAS.401.1552D,2010MNRAS.401.1043D}, composed of visually confirmed mergers from Galaxy Zoo Data Release 1 \citep[GZ DR1;][]{2011MNRAS.410..166L}. We complemented it with non-merging galaxies also from GZ DR1. The GZ DR1 project provided its volunteers with images of galaxies from a set of $\sim$900\,000 Sloan Digital Sky Survey Data Release 6 \citep[SDSS DR6;][]{2008ApJS..175..297A} galaxies. Through its interface, the users were asked questions related to the shape and visual properties, and the gathered answers resulted in a morphological classification. 

The mergers in \cite{2010MNRAS.401.1043D} are galaxies with $f_m>$ 0.4, where $f_m$ is the fraction of votes by GZ citizen scientists who thought the galaxy was a merger; the votes were weighted by their agreement rate with the majority opinion of the objects they viewed. These galaxies were then confirmed visually and their merging companions were selected manually by the authors in \cite{2010MNRAS.401.1043D}. The result was a public catalogue\footnote{\url{https://data.galaxyzoo.org/}} composed of merging galaxy pairs, including some multi-merger cases, with a total of 3$\,$003 merging systems in the redshift range [0.005,0.1].

To complete the dataset, we needed to include non-merging galaxies. As a preliminary non-merger set, we considered all GZ DR1 galaxies with $f_m<$ 0.2 \citep{2019A&A...626A..49P}. We built the training set to be class balanced, that is, each class that the NN learned to recognize was equally represented. Using a training set reproducing real abundances, the NN would become very good at finding non-mergers, but might not learn and confuse some of the less abundant mergers, thereby contradicting the purpose of this work. Besides having a class balanced set, we wanted to have a distribution of merging and non-merging galaxies similar in mass, so that they represent comparable populations in every property that is not related to the merging process. For example, if all mergers were significantly more massive than the non-mergers, the NN could be identifying the mass distribution rather than the merging state. While the $r$-band photometric magnitude is a good proxy for galactic mass \citep[e.g.][]{2018MNRAS.475..788M}, it is a detected flux that depends strongly on the distance. We decided to obtain one non-merging nearest neighbour for each merging galaxy in $r$-mag and spec-z simultaneously through a 2D euclidean distance.

We included a cut in low spec-z = 0.01 and another cut in high $r$-mag = 18.05; the former was because zero non-mergers could be found below that redshift, and the latter because the density of non-mergers decreased with $r$-mag. From the original 3$\,$003 primary mergers, four of them were found to show $f_m<$0.4, three even showing $f_m<$0.2, which meant that they appeared in both initial samples, so they were also removed.

The final dataset was reduced to 2$\,$930 mergers after all the restrictions were applied. Therefore, the 2$\,$930 non-mergers were matched and a final full dataset of 5$\,$860 galaxies was obtained. From this, we separated 5$\,$360 galaxies for training-validation purposes, and the remaining 500 for testing the trained NNs. In the next sections, we describe a variety of observational features of the dataset galaxies that we selected for NN inputs.

\subsection*{Magnitude types}
\label{subsect:mag}

The photometric magnitude of galaxies in SDSS is obtained for five bands: \textit{u}, \textit{g}, \textit{r}, \textit{i}, and \textit{z}. These magnitudes are calculated from the measured flux by an asinh magnitude function \citep{1999AJ....118.1406L}. Here we define the main magnitudes used for our experiment \citep{2002AJ....123..485S}:

\textbf{Point-spread-function flux:} is calculated by fitting the point spread function (PSF) function interpolated to the source position. The PSF with its spatial variation for each frame and band is measured by the SDSS's pipeline.

\textbf{Fibre flux:} the flux contained inside the aperture of the fibres of the SDSS spectrograph. Those fibres cover circular apertures of 3 arcseconds in diameter. In order to simulate better what the fibre sees in reality, the images are convolved with a 2-arcsecond seeing prior to the flux measurement. Appendix \ref{ap:1} describes how the magnitude and errors are derived from the aperture counts. 

\textbf{Petrosian magnitude:} the flux is calculated inside an aperture of radius $r_{\mathrm{P}}$. This radius is determined by forcing the inner flux to be a fixed factor -- the Petrosian ratio $R_{\mathrm{P}}$ -- of the mean flux on the circular annulus at the same $r_{\mathrm{P}}$. The aperture $r_{\mathrm{P}}$ set in the $r$ band is applied for the other four bands, so that the measurement is within a consistent band-independent aperture. 

\textbf{Exponential profile fit:} the flux is obtained from the fit of the galaxy's brightness distribution with an exponential profile, convolved with the PSF. The 2D exponential profile depends on the radius from the centre $r$ as $I(r)$, the surface brightness profile at $r$:
\begin{equation}
I(r) = I_0\, \mathrm{exp}\left[-1.68\frac{r}{r_0}\right] \, \, ,
\end{equation}
where $I_0$ is the flux at $r = r_0$, the half-light radius.

\textbf{De Vaucouleurs profile fit:}  fit with a De Vaucouleurs profile, convolved afterwords with the PSF. Its form is:
\begin{equation}
I(r) = I_0\, \mathrm{exp}\left[-7.67\left(\frac{r}{r_0}\right)^{\frac{1}{4}} \right] \, \,.
\end{equation}

\textbf{Model magnitude:} standard magnitude employed for SDSS data. It corresponds to a linear combination of the best fits with the exponential and the De Vaucouleurs profiles:
\begin{equation}
F_{\mathrm{model}} = frac_{DeV} F_{deV} + (1 - frac_{DeV}) F_{exp} \, \, ,
\end{equation}
where $frac_{DeV}$ is the linear combination factor that leads to the best possible fit of the profile combinations.

\section{Methodology}
\label{sect:Methodology}

An NN has combinations of layers, composed of basic mathematical nodes called neurons, connecting the input values to the model's final outputs. Each neuron weights the value of a previous point on the network -- which can be all the dataset inputs or all the neurons in the previous layer -- by its internal weight $\bm{g}$ and bias $\bm{b}$ parameters, performing a linear transformation of the type $\bm{f}(\bm{x}) = \bm{w}\cdot \bm{x} + \bm{b}$ that subsequently goes through a $g(\bm{f}(\bm{x}))$ non-linear activation function. $g(\bm{f}(\bm{x}))$ simulates a synaptic-like step in which the output is either zero or very small (no connection), or a larger number that allows the information to `progress' across the NN structure in some measure. The final output of the NN gives information about the input data such as their classification among a list of classes.

The NNs in this work were trained with known class labels (supervised learning), so that the neuron weights could be progressively modified and the models learnt to solve the task for which they had been built. Such a process may fall into a situation in which the NN explicitly memorizes the training dataset, a state known as overfitting. Neural network studies are mainly focused not only on how to increase the learning abilities, but also on how to reduce this overfitting and manage the generalization of the trained result. 

One of the controversies with NNs is that they tend to be treated as black boxes that somehow solve problems, even though it is not well understood how they manage them and why. This publication centres its attention not only on testing our NN models, but also on determining the information the NN finds in the dataset, and for that we made use of some other techniques such as dimensionality reduction, which we also explain in this section. The goal of the NN is not only to classify mergers, but also to learn their properties.

\subsection{Basics of our NN}

An NN is essentially defined by the layer-neuron architecture and the properties of the connections. We used dense layers, linking each of one layer's nodes to all those of the previous layer, creating subsequent, fully connected layers from input to output. In the intermediate step between layers, we applied batch normalization \citep{pmlr-v37-ioffe15} -- to normalize and shift the post-layer value array $\bm{l}$ to a mean $\bar{\bm{l}} = 0$ and variance Var($\bm{l}$) = 1 -- in order to make it faster and more stable. Moreover, we applied a dropout rate \citep{JMLR:v15:srivastava14a} for each layer, which consisted in setting some arbitrary percentage of the neurons to zero during each training step. This was done to allow all the neurons to be relevant in some way, forcing them to not rely on a higher influence from others.

The non-linear neuron activation function selected was the commonly used Rectified Linear Unit (ReLU \cite{10.5555/3104322.3104425}). The NN output is addressed as a two-class classification, also known as binary classification. The final result is given as a softmax probability value for the merger ($P_{\mathrm{mer}}$) and non-merger ($P_{\mathrm{nom}}$) classes. It fulfils $P_{\mathrm{mer}} + P_{\mathrm{nom}}$ = 1. The optimization method chosen was the Adam method \citep{2014arXiv1412.6980K}, characterized by a dynamic learning rate. The loss function used for optimizing the classifier was the TensorFlow's BinaryCrossentropy class. The layer layout, the dropout rate, and the initial learning rate selected are presented in Sect. \ref{subsect:NNarch}.

\subsection{Training}

The model was trained on the pre-selected combination of labelled objects detailed in Sect. \ref{sect:Dataset_selection}. Training datasets are commonly separated into three groups: a training set that is used for the optimization process, a validation set that is studied parallel to the training but without affecting the learning steps, and a test set that is only considered when the training is finished, to check the model's capabilities. During our NN learning, the training updates were done over a batch of 64 galaxies, randomly shuffled for every training epoch. At the same time, the validation set was used as control sample: updates on the neurons were externally saved only when there was an improvement in performance over the validation set. Specifically, we considered that the performance improved when both the validation loss decreased and the validation accuracy increased with respect to the last save. The test set was left unseen by the NN until the end, acting equivalently to applying the NN to a fully new group of objects.

For training the NN, we employed the k-fold cross-validation method \citep{10.2307/2984809, 5342427}. It consisted in separating the training dataset by shuffling objects randomly into $k$ equally sized groups. One full training was executed $k$ times so that every train-validation run had $k - 1$ groups forming the training-set and the remaining one as the validation-set, which was switched each time. As a result, we obtained $k$ trained NNs that could show the model instability and at the same time give a more reliable performance test. Moreover, if the learned parameters of the NN happened to be almost the same for all folds, an average of them could be used. For our 5$\,$360 training+validation galaxies, we decided to split them into $k$ = five folds, partly motivated to avoid extending the training time for too long. This adopted validation method will be referred to in the rest of the text as five-fold cross-validation. The advantage of applying our five-fold validation is that it allows us to compare NNs with different attributes or inputs by the mean and standard deviation of the validation peaks per fold.

\subsection{Input space normalization}
\label{subsect:input_space}

Inputs for NNs are generally normalized to facilitate the optimization process and simplify the numerical accuracy \citep{yu2005integrated}. For our inputs, we chose the min-max normalization. The original galaxy's feature array $X^{\mathrm{g}}$ was adapted to an [0,1] interval by applying the equation:

\begin{equation}
\centering
x^{\mathrm{g}} = \frac{X^{\mathrm{g}} - \mathrm{Min}(X^{\mathrm{g})}}{\mathrm{Max}(X^{\mathrm{g})} - \mathrm{Min}(X^{\mathrm{g})}} \, \, .
\label{eq:min_max}
\end{equation}

Different NNs were trained by separately using the different magnitude types described in Sect. \ref{subsect:mag}. However, the photometric information not only consisted of magnitudes but also of the measurement errors or the ten colour indexes -- derived by subtracting magnitudes in each band from one another. We considered multiple combinations of photometric information and labelled them as follows: an NN trained over an input space with only band magnitudes was labelled as B; with only colours was labelled as C; with bands and colours was labelled as BC; and with bands and colours and also with errors was labelled as BCE.

\subsubsection{Variations of the error normalization}

Additionally, for the BCE cases, we considered two more normalization formulas. This was due to the intrinsic relation of the errors $\sigma_B$ with the magnitudes, and thus normalizing them while ignoring this relation might lead to losing information. A more explicit possibility would be to relate the error normalization $\sigma_b$ to the measurement-error ratio.

While Eq. \ref{eq:min_max} was kept for bands and colours, two value corrections were used for $\sigma_b$, differing whether the post min-max band values $b$ were involved or not. The first case was obtained by the proportion of the original errors $\Sigma_B$ to the original band measures $B$, that is, the fractional error:

\begin{equation}
        \sigma_b = \frac{\Sigma_B}{B} \, \, .
\label{eq:prop}
\end{equation}
The other case was obtained by a relative normalization that maintains the same $\sigma_b/b$ ratio as the original error-magnitude $\Sigma_B/B$ one. This was obtained by solving an equality between the two fractions, leading to:

\begin{equation}
    \sigma_b = b \frac{\Sigma_B}{B} \, \, .
\label{eq:norm}
\end{equation}
Neural networks with their input normalized with Eqs. \ref{eq:prop} and \ref{eq:norm} are labelled as BCEp and BCEn, respectively.

\subsubsection{Min-max normalization of feature space, but not included fully}
\label{subsect:wo}

The multiple photometric parameters we combined to form the NN's input spaces had values varying by up to several orders of magnitude. When we min-max normalized them into the [0,1] interval, this sometimes distorted in some way the relation between them.  An illustrating example can be found in how the band magnitudes showed different normalized values between the input spaces of the BCE and the B versions of NNs. For the BCE case, the parameter with the maximum value was generally the \textit{u}-band magnitude, and the minimum was the error in some of the five bands. For the B case, while the maximum was the same, the minimum was the magnitude in some other band. Therefore, when applying Eq. \ref{eq:min_max} to the BCE space, the resulting input values for the band magnitudes were close to 1 and for the colours and errors close to 0. However, for the B case one could have a value 1 for the \textit{u}-band and 0 for the \textit{z}-band.

In some cases, to understand the role of specific parameters, it was necessary to isolate them. If one were interested in only studying the band's performance, min-max normalizing them alone would be the immediate option. Nonetheless, the values would by construction differ between the combined BCE and the isolated B input spaces, and making a comparison would not be easy. We attempted to mitigate this by getting the normalized values of a subset within a larger input set but discarding the non-interesting ones. As an example, to understand the role of the magnitude bands in the BCE space, we performed the same normalization as for the BCE NN but keeping only the band magnitude values and discarding the colours and errors after applying the min-max formula. We defined nomenclature of this type of input, using the bands' example, as follows: `bands as if in BCE' or`bands as if with colours and errors'.

\subsection{Formalism of NN parameters}
\label{subsect:formalism}
To quantify the success and performance of the NN, we define the following four classification groups: the mergers classified correctly are regarded as true positives (TPs); true negatives (TNs) are the non-mergers homologous; false negatives (FNs) are mergers mistakenly identified as non-mergers; and false positives (FPs) are non-mergers mistaken as mergers. Moreover, we consider accuracy as the ratio of correctly classified objects with respect to the whole set size:

\begin{equation}
    \mathrm{Accuracy} = \frac{\mathrm{TPs} + \mathrm{TNs}}{\mathrm{TPs} + \mathrm{TNs} + \mathrm{FPs} + \mathrm{FNs}} \,\, .
\end{equation}
For a single validation fold, the denominator would be the 1$\,$720 galaxies that compose it. The mergers' correctly classified rate is given as TPs/(TPs+FNs) and the non-mergers' correctly classified rate is given as TNs/(TNs+FPs).

\subsection{Dimensionality reduction}
\label{subsect:dim-red}

Dimensionality reduction techniques transform a high-dimensional set of data into a lower-dimensional representation. With the goal of simplifying a given problem or visualizing the data in a more adequate way, they attempt to maintain as much information of the original data as possible. In our case, we applied them with the purpose of visualizing the galaxies' distribution in 2D, reducing the original input dimensions.

We considered two different ways to do it, one of which was the principal component analysis \citep[PCA;][]{Hotelling1933AnalysisOA}, which is actually not a machine learning model but essentially a matrix diagonalization. The other way was the t-distributed Stochastic Neighbor Embedding \citep[t-SNE;][]{JMLR:v9:vandermaaten08a, vandermaaten2009dimensionality}, which is more oriented to resembling the original distribution.

\paragraph{Principal component analysis:} 
a linear method that performs a coordinate transformation of an N-dimensional dataset's feature space -- where each dimension corresponds to a data variable -- into a new N-dimensional orthonormal coordinate base. This new reference frame is composed of basis vectors called principal components. They arise as a consequence of diagonalizing the covariance matrix of the dataset and ordering the eigenvalues $\lambda_i$. The magnitude of $\lambda_i$ represents how high the variance is along the correspondent eigenvectors. Through PCA, one can obtain 2D or 3D plots with the projection of the data onto the directions with most feature variations.

\paragraph{t-distributed Stochastic Neighbor Embedding:} 
a machine learning algorithm that reduces an N-dimensional space into a 2D one while maintaining its statistical distribution as much as possible. This method calculates the relative probability of each pair of high-dimensional objects so that very similar or close points have high probabilities and very different ones have low probabilities. A similar probability distribution is initialized in 2D and, through minimizing the Kullback–Leibler divergence between both with respect to the point positions, an embedded map in 2D of the original space is produced.

\section{Results}

\subsection{Architecture selection}
\label{subsect:NNarch}
\begin{figure}%[h]
        \centering
        \includegraphics[width=\linewidth]{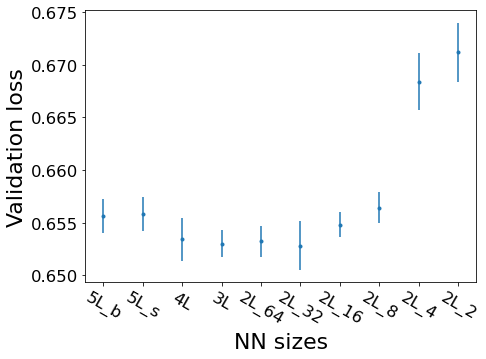}
        \caption{Validation loss for each tested NN architecture defined in Table \ref{tab:layouts}. It is calculated as the mean and standard error of the loss at the best validation update obtained in each of the five-fold validation cycles.}
        \label{fig:valLoss}
\end{figure}

In order to determine our definitive layer layout, we compared the NN architectures listed in Table \ref{tab:layouts} by their five-fold validation loss. The nomenclature prefix indicates the layer number as nL and the suffix refers to the size when required. Figure \ref{fig:valLoss} shows the mean loss and its standard error of the NNs sampled on the BC model magnitude input space. A relatively similar loss was obtained for all versions, except for the cases with too few neurons, 2L\_4 and 2L\_2. The longest NNs we tried combined up to five layers, but both 5L\_b and 5L\_s did not improve the performance. The 4L and 3L cases gave loss values as low as those of 2L\_64 and 2L\_32. Regarding instability, the architecture with the lowest variance was 2L\_16. For the definitive NN, we opted for 2L\_16 due to its convenient balance: a slightly worse but more stable loss when compared to 4L, 3L, 2L\_62, and 2L\_32, combined with a shorter computational time. We note that the five-fold galaxy distribution was not fixed but selected randomly for each architecture check.

\begin{table}%[h]
\begin{center}
        \centering  
        \caption{Architecture options considered for the NN layout, whose performances are given in Fig. \ref{fig:valLoss}. The Layout column gives the number of neurons per layer, separating each layer with a $+$.}
        \label{tab:layouts}
        \begin{tabular}{l  c } 
                NN name & Layout \\ \hline \hline
                \footnotesize 5L\_b & \footnotesize 128+256+512+256+128  \\  
                \footnotesize 5L\_s & \footnotesize 64+256+512+256+64 \\ 
                \footnotesize 4L & \footnotesize 128+256+256+128        \\
                \footnotesize 3L & \footnotesize 32+128+32 \\
                \footnotesize 2L\_64 & \footnotesize 64+64 \\   
                \footnotesize 2L\_32 & \footnotesize 32+32 \\
                \footnotesize 2L\_16 & \footnotesize 16+16 \\ 
                \footnotesize 2L\_8 & \footnotesize 8+8 \\
                \footnotesize 2L\_4 & \footnotesize 4+4 \\
                \footnotesize 2L\_2 & \footnotesize 2+2 \\ \hline \hline
        \end{tabular}
\end{center}
\end{table}

\begin{figure}%[h]
        \centering
        \includegraphics[width=\linewidth]{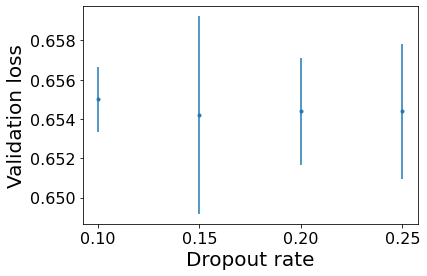}
        \caption{Validation loss for each dropout rate chosen. The value is again the mean of the five-fold validation cases and the error bars come from the standard error among the five folds.}
        \label{fig:dropout}
\end{figure}

The other tested NN parameters were the initial learning rate of the Adam optimizer and the dropout rate. The former performed quite badly when significantly diverted from 5$\times$10$^{-5}$. Regarding the latter, the validation losses of the considered variations are shown in Fig. \ref{fig:dropout}. Following the argument for selecting 2L\_16, a 0.1 rate would be a more adequate option as it shows a larger loss and smaller standard error. However, we decided to keep the 0.2 rate because the error bar for 0.1 covers only the upper -- and worse -- part of the interval that 0.2 covers.

\subsection{Input magnitude variations}

We established the BC model magnitude as the initial reference NN for comparisons because it more closely represents the real galaxies' brightness. This input is a 15D space combining the five photometric bands plus the resulting ten colours, normalized in the range [0,1]. It led to an accuracy of 68.90 $\pm$ 0.72\%, as shown in Table \ref{tab:ref}. This accuracy implies that we found a stable classifier capable of correctly identifying a substantial amount of objects. Moreover, it is encouraging how around $60\%$ of the mergers were correctly classified, given such a simple input space.

\begin{table}%[h]
    \centering
    \caption{Reference NN training parameters as defined in Sect. \ref{subsect:formalism}: the bands plus colours model magnitude case. They imply the NN has the potential of figuring out the classification rules.}
    \begin{tabular}{c c}
    \hline \hline
      \footnotesize Accuracy &  \footnotesize 68.90 $\pm$ 0.72 \% \\
      \footnotesize TPs & \footnotesize 61.65 $\pm$ 0.77 \% \\
        
      \footnotesize TNs &  \footnotesize 75.49  $\pm$ 0.92 \% \\
    \hline \hline
    \end{tabular}
    \label{tab:ref}
  \end{table}

\begin{figure}
    \centering
    \includegraphics[width=\linewidth]{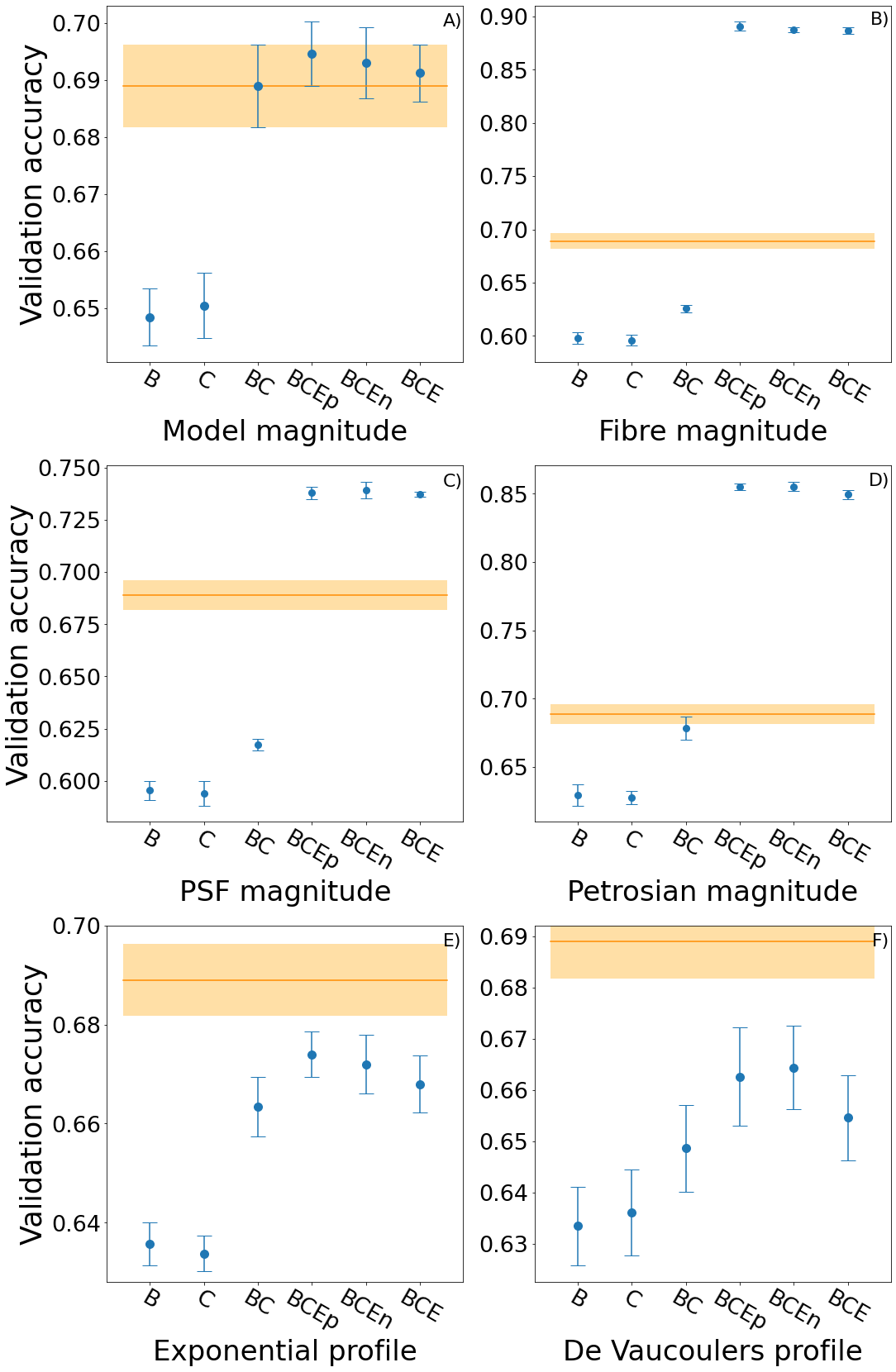}
    \caption{Six-panels plot showing the mean validation peak accuracy of the different input variations for each magnitude type. For each magnitude defined in Sect. \ref{subsect:mag}, we provide several variations: B, which corresponds to the five band values; C, the ten colours obtained from the five bands; BC, a 15-dimensional space combining bands and colours; and BCE, a 20-dimensional space that adds the magnitude errors to the BC cases. All four of these variations follow the min-max normalization defined in Eq. \ref{eq:min_max}. Additionally, we show the BCEp and BCEn sets, for which bands and colours were min-max normalized separately to the errors, obtained with Eqs. \ref{eq:prop} and \ref{eq:norm} respectively. The distribution of galaxies among the five validation folds is fixed to be the same. Panel A) corresponds to the model magnitude type, B) to the fibre magnitude, C) to the PSF magnitude, D) to the Petrosian magnitude, E) to the exponential magnitude, and F) to the De Vaucouleurs one.}
    \label{fig:all_res}
\end{figure}

Figure \ref{fig:all_res} shows the resulting accuracy of the NN applied over all six types of magnitudes -- as defined in Sect. \ref{subsect:mag} -- and the input variations -- as defined in Sect. \ref{subsect:input_space}. The horizontal orange line is the reference NN accuracy, and the shaded area is its error. Each panel in Fig. \ref{fig:all_res} confirms how the BC magnitude inputs lead to a significant increase in the accuracy with respect to the separated B and C cases. Both bands and colours generate consistent accuracies in all magnitude classes because they essentially contain the same information: one colour index is nothing more than a linear combination of two-band magnitudes, the ratio of two-band fluxes. Including the two of them at the same time seems to facilitate the NNs' performance, and therefore seems to be a way to improve the model. This pattern is found independently of the magnitude type.  

Introducing the errors in the feature space has an influence that highly depends on the magnitude type. On the one hand, for the model magnitude, the exponential, and the De Vaucouleurs profiles, the inclusion of errors produces an only slight increase in the accuracy. On the other hand, for the PSF, Petrosian, and fibre magnitudes, the accuracy increases highly, reaching almost 90\% for the BCEp fibre case.

Regarding the error normalization options, both the BCEp (Eq. \ref{eq:prop}) and BCEn (Eq. \ref{eq:norm}) cases have slightly better accuracy than BCE cases, while being consistent within the standard error. The BCEn cases are generally slightly better than the BCEn ones. This is one indication that suggests that the NN is independent of the normalization method.

\subsection{Fibre errors mixed with other data}

The surprising success from the fibre BCE case deserved a deeper look: not only did it reach a high accuracy, but it also outperformed the reference input -- the BC model magnitude. We narrowed down the search of what the main information source was within fibre BCE by combining the bands, colours, and errors of both model and fibre magnitudes in different ways.

Table \ref{tab:mix} shows the validation mean accuracy for all the relevant combinations, separated into five blocks. The first two blocks indicate that the model B, C, or BC cases combined with model E retain an accuracy of around 67 \%, but if the model errors are swapped with fibre errors, then the NN achieves high values, similar to those of the BCE fibre. The third and fourth blocks confirm the same results, but applying fibre B, C, and BC cases. The last block shows the results for model E or fibre E alone. The importance of fibre E is therefore demonstrated not only by how these errors enhance the accuracy when they accompany any magnitude type but also by their performance when used as a 5D input.

\begin{table}%[h!]
\centering
\caption{Validation mean accuracy and standard error for all the relevant combinations, separated into five blocks. Each input space undergoes a min-max normalization. The table's first two blocks combine the model B, C, and BC cases with either model E or fibre E, respectively; the third and fourth blocks do the same but for the fibre B, C, or BC cases; and the last block shows what happens for model E or fibre E alone. As in Fig. \ref{fig:all_res}, the source distribution among the five validation folds is fixed to be the same.}
\begin{tabular}{c c}
        Input space & Accuracy \\
        \hline \hline
        \footnotesize{model B + model E} & \footnotesize{67.14 $\pm$ 0.31} \% \\
        \footnotesize{model C + model E} & \footnotesize{67.24 $\pm$ 0.64} \% \\
        \footnotesize{model BC + model E} & \footnotesize{69.12 $\pm$ 0.50} \% \\
        \hline
        \footnotesize{model B + fibre E} & \footnotesize{77.08 $\pm$  0.73} \% \\
        \footnotesize{model C + fibre E} & \footnotesize{88.30 $\pm$ 0.45} \% \\
        \footnotesize{model BC + fieer E} & \footnotesize{87.40 $\pm$ 0.72} \% \\
        \hline
        \footnotesize{fibre B + model E} & \footnotesize{62.56 $\pm$  0.51} \% \\
        \footnotesize{fibre C + model E} & \footnotesize{62.40 $\pm$ 
        0.45} \% \\
        \footnotesize{fibre BC + model E} & \footnotesize{62.80 $\pm$ 0.55} \% \\
        \hline
        \footnotesize{fibre B + fibre E} & \footnotesize{77.14 $\pm$ 0.60} \% \\
        \footnotesize{fibre C + fibre E} & \footnotesize{88.66 $\pm$ 0.37} \% \\   
        \footnotesize{fibre BC + fibre E} &  \footnotesize{87.84 $\pm$ 0.42} \% \\
        \hline
        \footnotesize{model E} & \footnotesize{59.06 $\pm$ 0.76} \% \\
        \footnotesize{fibre E} & \footnotesize{83.76 $\pm$ 0.32} \% \\
        \hline
        \hline
\end{tabular}
\label{tab:mix}
\end{table}

\subsection{Fibre errors components}
\label{subsect:fiber_components}

According to the NN results presented up to now, one can just use the fibre magnitude error and get a correct classification with an accuracy of $\sim$84\%. Such an achievement does not seem intuitively justified solely by the properties of the input data. The next step was then to understand how fibre E values were calculated.

According to the SDSS documentation, the fibre magnitudes are simple aperture magnitudes, meaning we can replicate them with the original astronomical frames and the appropriate calibration data. The whole process of reproducing the fibre magnitudes and errors is detailed in Appendix \ref{ap:1}. We obtained the errors from the fibre aperture counts in the fpObjc file downloaded from the SDSS repository, sampled in each band on a group of ten galaxies. We justify the updated version of the formula that calculates the error from the initial counts (Eq. \ref{eq:new_CE}).

According to Eq. \ref{eq:new_CE}, the aperture error calculation has four particular inputs per band, which are: the digital unit count in the aperture region \code{counts}, the error in the CCD camera's dark current \code{dark variance}, the sky background estimation in the source's centre \code{sky}, and its error \code{skyErr}. We retrieved the four inputs per band for each galaxy in our kfold validation sample. We trained several NNs, noted in Table \ref{tab:fE_input}, considering first all sets of variables and then excluding each one at a time. It arose that the main source of accuracy was \code{skyErr}, as expressed in the accuracy from fibre E inputs without \code{skyErr}. Furthermore, running an NN only with \code{skyErr} showed a very similar accuracy to that of fibre E alone, implying they contain similar information.

Moreover, every training in which \code{skyErr} was accompanied with other features achieved an accuracy better than 90\%. The best NN is for \code{skyErr} and \code{dark variance} together. Nonetheless, the dark current error does not seem to be related to the galactic properties at all, as it is a property of the CCD camera. This led to the last result, checking the dependence of the NN accuracies on the input normalization.

\begin{table}
\centering
    \caption{Validation peak mean and error for the NN input spaces using different variables that combine to make up fibre E. The first row considers all inputs shown in Eq. \ref{eq:new_CE}, followed by the results of excluding one variable at a time. The next row shows the case of only \code{skyErr}, and the final three rows come from combining \code{skyErr} with every other variable.}
        \begin{tabular}{c c}
                Input space & Accuracy \\
                \hline  \hline
                \footnotesize{fibre E -- all-input-set} & \footnotesize{91.48 $\pm$ 0.31} \% \\
                
                \hline
                \footnotesize{fibre E -- without \code{counts}} & \footnotesize{91.62 $\pm$ 0.25} \% \\
                \footnotesize{fibre E -- without \code{dark variance}} & \footnotesize{91.48 $\pm$ 0.23} \% \\
                \footnotesize{fibre E -- without \code{sky}} & \footnotesize{91.98 $\pm$ 0.27} \% \\
                \footnotesize{fibre E -- without \code{skyErr}} & \footnotesize{60.62 $\pm$ 0.85} \% \\
                \hline
                \footnotesize{\code{skyErr}  -- only} & \footnotesize{83.30 $\pm$ 0.38} \% \\
                \hline
                \footnotesize{\code{skyErr} + \code{counts}} & \footnotesize{91.66 $\pm$ 0.34} \% \\
                \footnotesize{\code{skyErr} + \code{dark variance}} & \footnotesize{92.04 $\pm$ 0.26} \% \\
                \footnotesize{\code{skyErr} + \code{sky}} & \footnotesize{90.90 $\pm$ 0.21} \% \\
                \hline \hline
\end{tabular}
\label{tab:fE_input}
\end{table}

\subsection{Normalization dependence}
\label{subsect:norm-dependance}

The high accuracy our NN obtained from the \code{skyErr} and fibre E inputs varied with the companion features. We wanted to see if this behaviour was because of the normalization applied. The first test was to expand the min-max normalization interval from [0,1] to [0,2] for some selected cases. In Table \ref{tab:main_NNsx2} the accuracies of the four most relevant inputs are compared between both versions. Only for the fibre E all-input set is there a relatively significant difference, leading us to conclude that the min-max resulting normalized interval is not a critical choice.

\begin{table}
    \centering
    \caption{Main project NNs but with the min-max normalization set to an [0,2] interval. Little difference can be found from the previous case.}
    \begin{tabular}{c c c}
      Input Space & \makecell{Accuracy \\ min-max [0,2]}  & \makecell{Accuracy \\ min-max [0,1]} \\
      \hline \hline
      \footnotesize Reference NN & \footnotesize 68.82 $\pm$ 0.57 \%  & \footnotesize  68.90 $\pm$ 0.72 \%\\
        
          \footnotesize Fibre BCE & \footnotesize 88.64 $\pm$ 0.36 \% & \footnotesize 88.68 $\pm$ 0.31 \% \\
          
          \footnotesize Fibre E -- all-input-set & \footnotesize  90.82 $\pm$ 0.14 \% & \footnotesize 91.48 $\pm$ 0.31 \% \\
          \footnotesize \code{skyErr} -- only &  \footnotesize 83.52 $\pm$ 0.38 \% & \footnotesize 83.30 $\pm$ 0.38 \% \\
          
          \hline \hline
          
    \end{tabular}
    \label{tab:main_NNsx2}
  \end{table}

The next step was to apply the normalization method introduced in Sect. \ref{subsect:wo} to different input spaces that included either \code{skyErr} or fibre E. Table \ref{tab:norm_variations} shows the accuracy of the NNs we built using the input \code{skyErr} or fibre E subsets, compared to the complete spaces from which we isolated them. When \code{SkyErr} is normalized with \code{dark variance}, or with \code{counts} and \code{dark variance}, or with all other error inputs -- \code{counts}, \code{sky,} and \code{dark variance} -- the accuracy shows little dependence on the companions' presence or absence. For the \code{skyErr} without normalization, the accuracy is 90.88 \%, which is also high. Fibre E is shown to be related to the companion for the fibre BCE and fibre CE cases, but not for the BCEp (fractional errors) or the BCEn case. However, the errors in the BCEp and BCEn cases are explicitly obtained from the bands, so it could be argued that the NN does get that information from the resulting error inputs. The pre-normalized fibre E shows better values than its min-max, but it deviates from the other cases more than \code{skyErr} does.

It can be interpreted that fibre E does benefit from being together with the magnitudes in specific combinations, but that \code{skyErr} is sufficient by itself. Nonetheless, the other conclusion is that \code{skyErr} depends on the normalization to provide the high accuracy observed. Because the best NN result comes from the \code{skyErr} as if with \code{dark variance} case, a parameter unrelated to the galaxies, and the resulting accuracy is close to that of the pre-normalized \code{skyErr} input, we can consider that the role of the \code{dark variance} is simply to adapt the \code{skyErr} numerical representation for the NN to better identify the properties of the mergers. Moreover, the last row with the sky error in logarithmic values achieves an even better result, supporting the finding that the \code{skyErr} is a good merger proxy on its own. The NN for this best case with the saved weights can be found on \code{GitHub}\footnote{\url{https://github.com/LuisEduSuelves/NN16_skyErr-log}}.

\begin{table}
    \centering
    \caption{Neural network performance for input spaces in which either \code{skyErr} or fibre E has been normalized with some other parameters that were not included as inputs (see Sect. \ref{subsect:wo}). The last column gives the accuracy for the input space prior to the applied modification. For the \code{skyErr} and fibre E pre-normalized spaces (rows 4 and 9), the last column compares their respective 5D isolated min-max value.}
    \begin{tabular}{c c c }
      Input space & Accuracy & \makecell{Previous \\ accuracy} \\
    \hline \hline
    \footnotesize \makecell{\code{skyErr} \\ as if with  \\ \code{dark variance}} & \footnotesize 92.10 $\pm$ 0.28 \% & \footnotesize 92.04 $\pm$ 0.26 \% \\
    \hline
    \footnotesize \makecell{\code{skyErr} \\ as if with \\ \code{counts} \& \\ \code{dark variance}} & \footnotesize 92.02 $\pm$  0.21 \% & \footnotesize 91.98 $\pm$ 0.27 \%   \\
    \hline
    \footnotesize \makecell{\code{skyErr} \\ as if in \\ fibre E -- all-input-set} & \footnotesize  91.02 $\pm$ 0.32 \% & \footnotesize 91.48 $\pm$ 0.31 \% \\
    \hline
    \hline
    \footnotesize \makecell{\code{skyErr} \\ pre-normalized} & \footnotesize 90.88 $\pm$ 0.25 \%  & \footnotesize 83.30 $\pm$ 0.38 \% \\
    \hline
    \hline
    \footnotesize \makecell{fibre E \\ as if in BCE} & \footnotesize 79.72 $\pm$  0.32\% & \footnotesize  88.68 $\pm$ 0.31 \% \\
    \hline
    \footnotesize \makecell{fibre fractional \\ errors} & \footnotesize 88.60 $\pm$ 0.55 \%  & \footnotesize 89.06 $\pm$ 0.42 \% \\
    \hline
    \footnotesize \makecell{fibre E \\ as if in BCEn} & \footnotesize 88.40 $\pm$ 0.33 \%  & \footnotesize  88.76 $\pm$ 0.25 \% \\
    \hline
    \footnotesize \makecell{fibre E \\ as if in CE} & \footnotesize 77.66 $\pm$ 0.66 \%  &  \footnotesize 88.66 $\pm$ 0.37 \% \\
    \hline
    \hline
    \footnotesize \makecell{fibre E \\ pre-normalized} & \footnotesize 86.74 $\pm$ 0.40 \% & \footnotesize 83.76 $\pm$ 0.32 \% \\
    \hline
    \hline
    \footnotesize\makecell{\code{skyErr} pre-normalized \\  logarithmic scale} & \footnotesize 92.64 $\pm$ 0.15 \% & ... \\
    \hline
    \hline
    \end{tabular}
    \label{tab:norm_variations}
  \end{table}

\section{Discussion}

The accuracies obtained from the most relevant input spaces are presented in Table \ref{tab:main_NNs}, together with their performance on the test set applying the saved weights. These results demonstrate that the NN has successfully classified galaxy mergers by making use of photometry, and that we have found the sky background error to be the source of the best method. Such a calibration parameter potentially points to the importance of differential image analysis, which, to our knowledge, had not been considered as a key method for galaxy merger identification until now. Therefore, this discussion will attempt to justify the advantages of our method. In Sect. \ref{subsect:reproducibility}, we address its reproducibility and its potential use both in SDSS and in other surveys. Then,  in the rest of the section, we study the distribution of galaxies in the five-band \code{skyErr} space using the min-max normalization as if with \code{dark variance}, which is the next-to-best accuracy found. For that, we show dimensionality reduction and feature space distributions. To summarize, we infer why the logarithmic sky error should work even better as input space, showing that a simple 2D boundary can be as effective as the NN. Finally, we justify why the sky background error contains information of merging processes.

\begin{table}
    \centering
    \caption{Accuracies for the central NN input spaces of the project. The table also gives the mean accuracy over the test set, calculated using the weights at the peak validation in each fold.}
    \begin{tabular}{c c}
      \footnotesize Input space & \footnotesize Accuracy \\
    \hline \hline
      \footnotesize Reference NN &  \footnotesize 68.90 $\pm$ 0.72 \% \\
          
          \footnotesize Test Reference NN & \footnotesize 69.72 $\pm$ 0.36 \% \\
        \hline
          \footnotesize Fibre BCE & \footnotesize 88.68 $\pm$ 0.31 \% \\
          \footnotesize Test Fibre BCE & \footnotesize 89.60 $\pm$ 0.24 \% \\
        \hline
          \footnotesize Fibre E -- inputs & \footnotesize  91.48 $\pm$ 0.31 \% \\
          \footnotesize Test Fibre E --inputs & \footnotesize 91.20 $\pm$ 0.35 \% \\
        \hline
          \footnotesize \code{skyErr} &  \footnotesize 83.30 $\pm$ 0.38 \% \\
          
          \footnotesize Test \code{skyErr} & \footnotesize 79.56 $\pm$ 0.10 \% \\
        \hline
          \footnotesize \makecell{\code{skyErr} min-max as if \\ with  \code{dark variance}} & \footnotesize 92.10 $\pm$ 0.28 \% \\ 
          
          \footnotesize \makecell{Test \code{skyErr} min-max as if \\ with \code{dark variance}}  & \footnotesize  90.92 $\pm$ 0.20 \% \\
        \hline
          \footnotesize\makecell{\code{skyErr} pre-normalized \\ in logarithmic scale} & \footnotesize 92.64 $\pm$ 0.15 \%. \\
          \footnotesize\makecell{Test \code{skyErr} pre-normalized \\ logarithmic scale} & \footnotesize 92.36 $\pm$ 0.21 \%. \\
        \hline \hline
    \end{tabular}
    \label{tab:main_NNs}
\end{table}

\subsection{Reproducibility of the model}
\label{subsect:reproducibility}

The step-by-step measurement of the sky background error\footnote{\url{http://classic.sdss.org/dr6/algorithms/sky.html}} should be easy to reproduce in any other optical survey because the measurements are very generic. Nothing the pipeline does should be unusual for another astronomical survey and there is no dependence on the SDSS specifications in any step. It may be that the cut-out box size where the local background is estimated should be adapted to different pixel sizes if necessary. We cannot foresee beforehand if any other SDSS specification might have some influence, but it does not seem the case at the present stage. Regarding the training sample, its redshift and mass distribution define the range in which the NN is, in principle, effective. To what extent the NN could be applied to sources outside this data will be addressed in a forthcoming work.

\subsection{Sky error properties found by the NN. Case \code{skyErr} as if with \code{dark variance}}

\begin{figure}
        \centering
        \includegraphics[width=\linewidth]{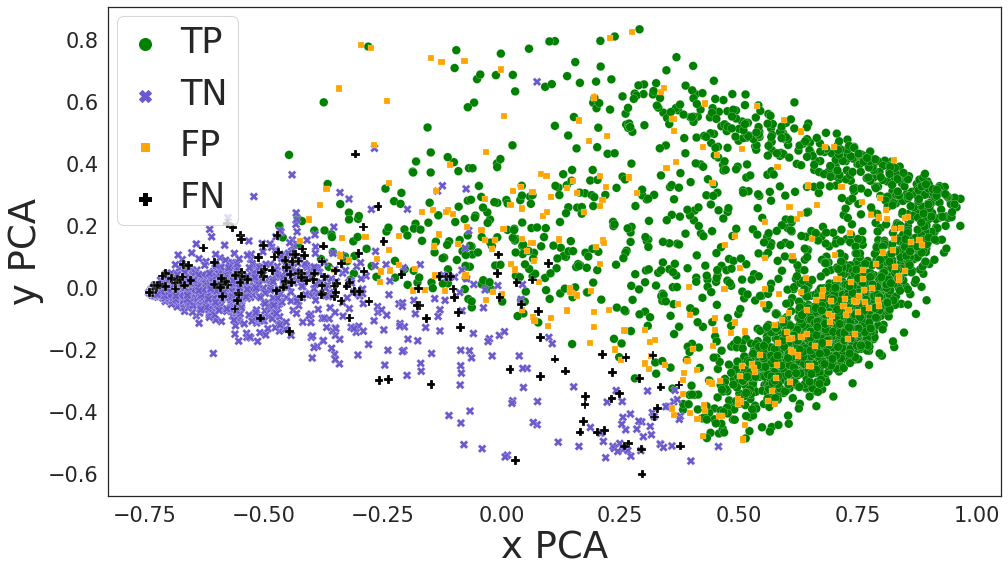}
        \caption{2D embedding using PCA. Classification results from the weights saved at the validation peak of the first of the five folds: TP galaxies are shown by green circles, TNs by blue crosses, FNs by black `x's, and FPs are in orange. The axes are the first and second principal coordinates. This colour scheme will be repeated for all the plots in the rest of the text.}
        \label{fig:PCA_skyErr-wo-dV}
\end{figure}

\begin{figure}
        \centering
        \includegraphics[width=\linewidth]{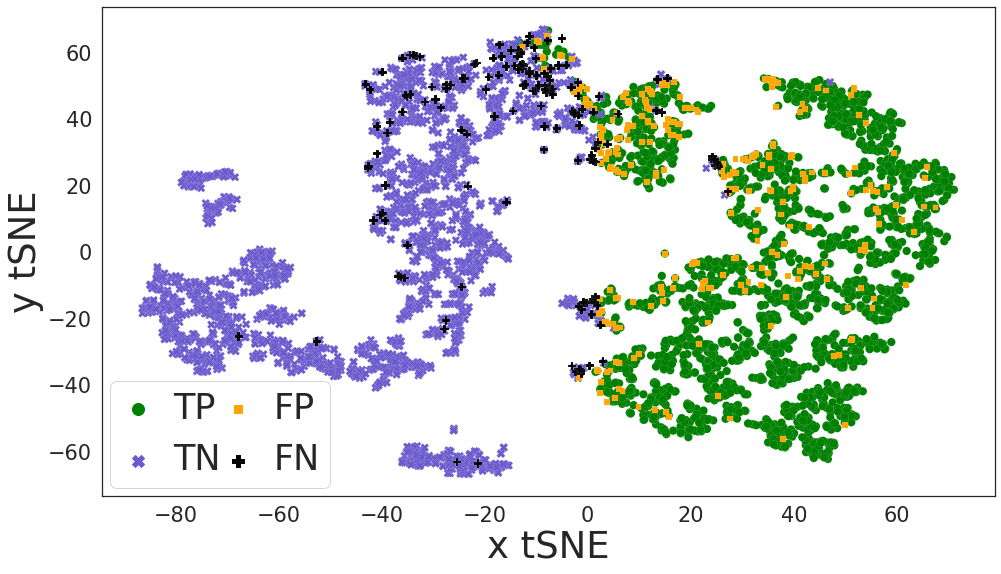}
        \caption{2D embedding using tSNE, with the same classification scheme as for Fig. \ref{fig:PCA_skyErr-wo-dV}. The axes are simply the two tSNE dimensions.}
        \label{fig:tSNE_skyErr-wo-dV}
\end{figure}

Figures \ref{fig:PCA_skyErr-wo-dV} and \ref{fig:tSNE_skyErr-wo-dV}, show, respectively, the PCA and tSNE methods -- Section \ref{subsect:dim-red} -- applied to the \code{skyErr} input space set normalized with \code{dark variance}. Each data point corresponds to a galaxy in the 2D embedding and the colour depends on the classification type (Sect. \ref{subsect:formalism}). Because t-SNE is highly dependent on the initial distribution, we initialized the model based on the data's PCA, a functionality available in the \code{python sklearn} package that we implemented.

Through the PCA plot, the main locations for mergers and non-mergers, where TPs and TNs are denser, can be seen in opposite corners of the rhomboid shape. The TPs are to the right and the TNs are to the left, while in the intermediate area, the plot is less dense and more FNs and FPs arise mixed in between. Some FNs or FPs appear also in the green and blue dense areas, respectively. This is more frequent for FPs, which indicates that the \code{skyErr} method still does not define an unmistakable distinction between mergers and non-mergers.

The tSNE method leads to very similar conclusions but from a more defined shape with a more uniform density. The TP versus TN separation is delimited in a clear way. Again, some FPs and FNs are dotting the TP and TN areas. The FNs appear rarely in the blue region; they arise mostly in the TN edges that can be found even intersecting green regions, such as near [0,-10] or [20,30].

\begin{figure}
        \centering
        \includegraphics[width=\linewidth]{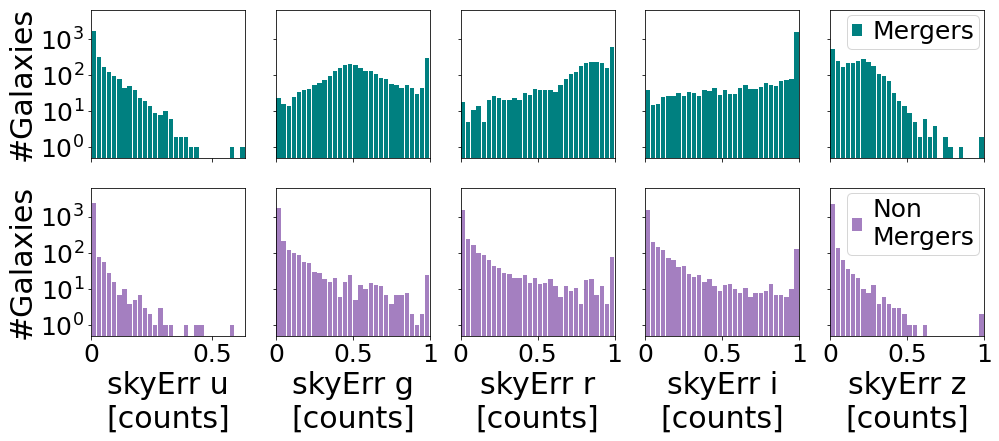}
        \caption{Sky background error histogram panels for the five bands. Galaxies labelled as mergers are in blue and non-mergers are in light red. The values were normalized with the dark current variance.}
        \label{fig:hist-label}
\end{figure}

\begin{figure}
        \centering
        \includegraphics[width=\linewidth]{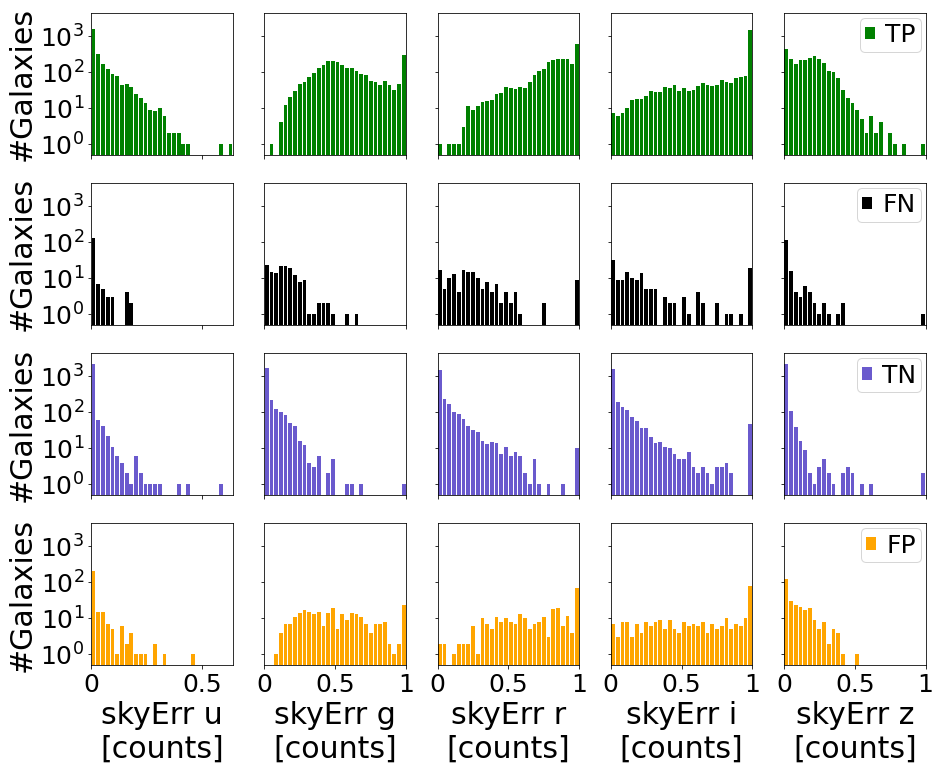}
        \caption{Distribution of the same variables as in Fig. \ref{fig:hist-label}, but split into the four classification types.}
        \label{fig:hist-class}
\end{figure}

Another approach to investigate the data is to create histograms of the inputs. For that, we show the histogram per label in Fig. \ref{fig:hist-label} and the histogram per class in Fig. \ref{fig:hist-class}. Figure \ref{fig:hist-label} shows that for mergers, the distribution in the $u$ and $z$ bands has a less steep profile than for non-mergers. The mergers show one peak around 0.2 in $z$ that disappears for the non-mergers. It translates into a defining characteristic, as the distribution is maintained in the TPs in Fig. \ref{fig:hist-class}. In contrast, the $g$, $r$, and $i$ bands present distributions that differ more between labels. For non-mergers, the three of them peak near 0 and decrease in number towards larger normalized errors. For mergers, there is a peak that translates from central values in $g$ to progressively larger ones in the other two bands. An immediate interpretation is that the majority of mergers present an intermediate relative sky error in $g$, a relatively high error in $r$, or the highest possible value 1, for a single galaxy after min-max in $i$. Overall, the TPs distribution in Fig. \ref{fig:hist-class} resembles the mergers' one. The TNs simply show a steep slope similar to that in $u$ and $z$, with the FNs again also flat. The FPs histograms are quite uniform in comparison to the others. It seems that the FPs cover the part of the distribution of non-mergers that goes missing in the TN distributions. The main signs the NN is identifying become evident when comparing TPs and TNs in $g$, $r$, and $i$ bands. The mergers have high \code{skyErr} values but the non-mergers have low ones. The FN and FP profiles indicate that this is neither sufficient nor clearly defined, as the dimensionality reductions were illustrating.

\begin{figure}
        \centering
        \includegraphics[width=\linewidth]{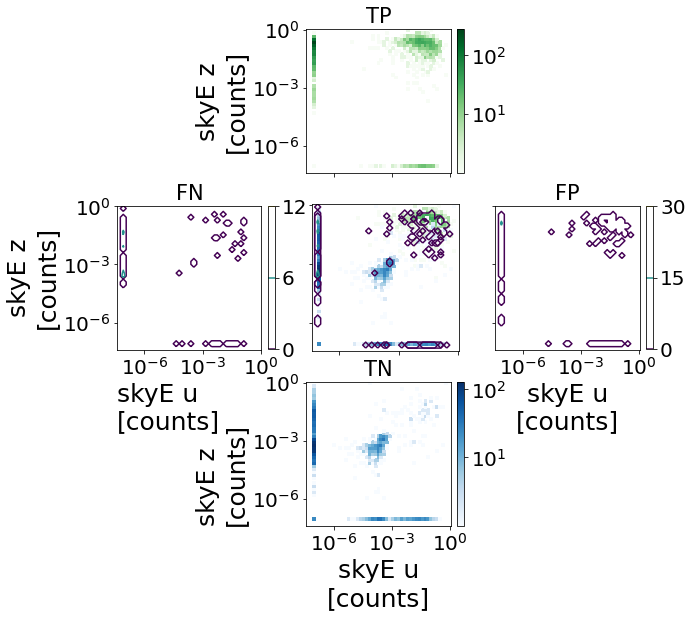}
        \caption{Distribution of galaxies in the 2D histograms of the TPs (in green, separated above) and TNs (blue, below), and the contour plots of the FNs (left), FPs (right), and of all galaxies (centre) for \code{skyErr} in the $u$-band vs  the $z$-band plane. The 2D histograms show logarithmic colour-bars and the axes are in logarithmic scale. To avoid undefined values for the galaxies with post-normalization features equal to zero, a constant value of $10^{-7}$ was added to these. Consequently, they appear as vertical and horizontal lines at the bottom and left sides of each panel. This allows us to see what happens with those. It should be noted that some TNs are in $10^{-7}$ for each band, meaning the pre-normalized \code{skyErr} in both bands was exactly the same for those galaxies.}
        \label{fig:2D_u-z}
\end{figure}

\begin{figure}
        \centering
        \includegraphics[width=\linewidth]{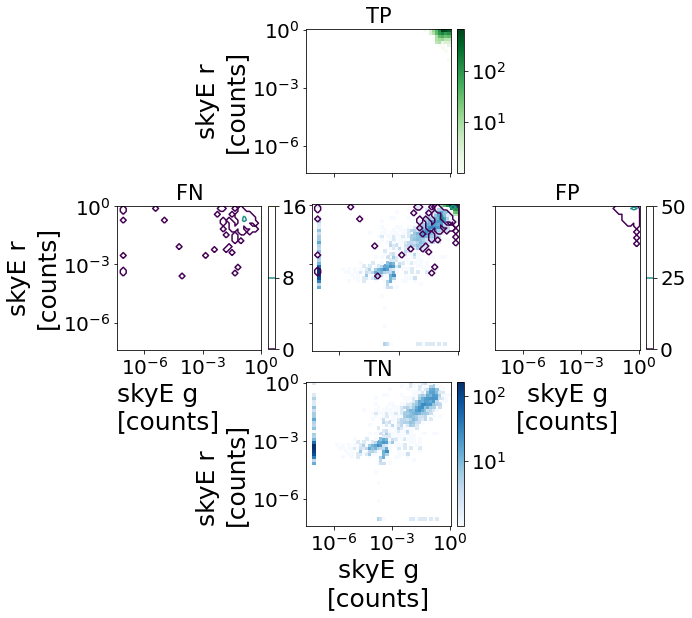}
        \caption{Same panels as in Fig. \ref{fig:2D_u-z}, but this time for bands $g$ and $r$.}
        \label{fig:2D_g-r}
\end{figure}

Figures \ref{fig:2D_u-z} and \ref{fig:2D_g-r} show the 2D histograms of the TPs and FPs, and the contour plots of the FNs and TPs of \code{skyErr} for the $u$ versus $z$ and the $g$ versus $r$ bands, respectively. The TPs and TNs in the first image are clearly separated, although some TNs can be seen in the upper right area, where the TPs are mostly located. The FP and FN contour plots show the area where the confusing galaxies are located. Similar to the patterns appearing in the dimensionality reduction figures, the FPs are mostly around the TP area and the FNs appear both in the intermediate region and near the TNs. For $g$-$r$, the location of the TPs is mostly in the upper right corner, near to a value of 1. Analogously to the first image, the TP and TN areas are clearly separated, the FPs are located mostly in the same region as the TPs, and the FNs appear both in the TN region and in the intermediate TP-TN area. Therefore, the properties in the 1D histograms can be seen translated into a 2D representation and the dimensionality reduction patterns are present.

\subsection{Pre-normalized \code{skyErr}}
\label{subsect:pre-norm_skyerr}

\begin{figure}%[h]
        \centering
        \includegraphics[width=\linewidth]{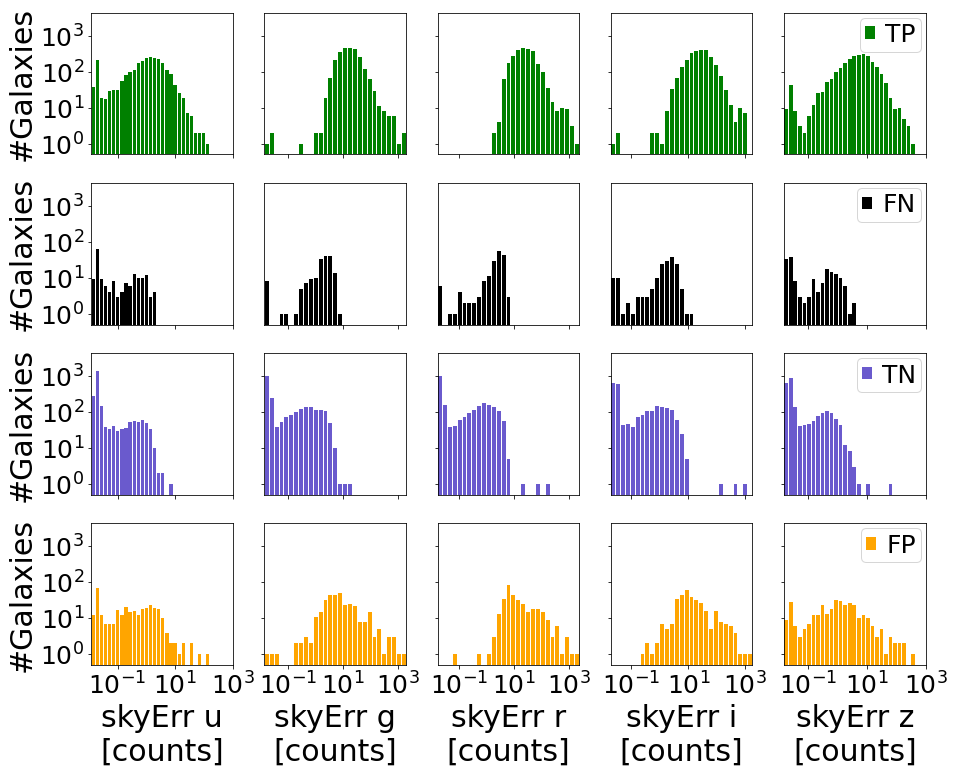}
        \caption{Sky background error original histograms in logarithmic bins and bin widths.}
        \label{fig:hist_skyErr_pre-norm-log}
\end{figure}

To better understand the pre-normalized \code{skyErr} features, we created individual histograms as in Fig. \ref{fig:hist_skyErr_pre-norm-log}, but using logarithmic bins to enable a better visualization. The separation between the distributions depending on the classification type is even more pronounced here than in Fig \ref{fig:hist-class}. For bands $g$, $r$, and $i$, the TPs and FPs are located in the upper half of the data, and the TNs and FNs are in the lower half. This strong separation is not seen in the $u$ and $z$ bands. These patterns are analogous to those described in the previous section, but are even more explicit. We created one last input space with the logarithm of the \code{skyErr} to check if the NN is aware of this difference in the data's presentation. This provided an accuracy of 92.64 $\pm$ 0.15 \%, the best result obtained in this project and shown in the last row in Tables \ref{tab:norm_variations} and \ref{tab:main_NNs}. The accuracy is 2\% better than that corresponding to the linear pre-normalized \code{skyErr}, confirming the importance of the normalization. Therefore, we can conclude that the normalization of the data plays a crucial role in the success of our model.

\begin{figure}
        \centering
        \includegraphics[width=\linewidth]{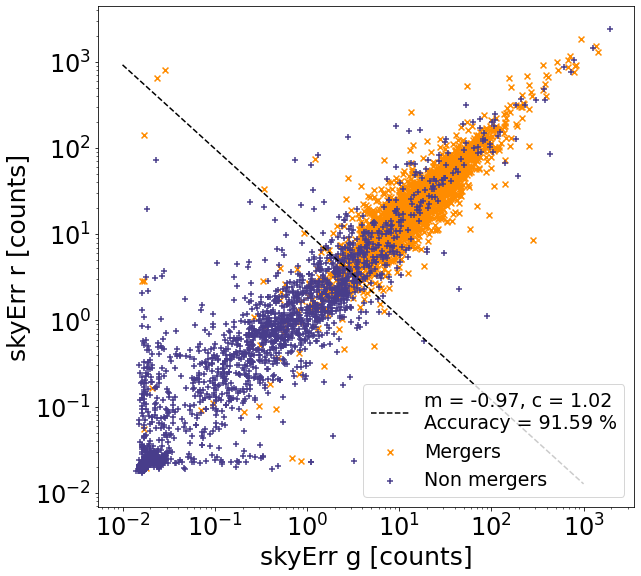}
        \caption{Distribution of galaxies in the 2D plane of \code{skyErr} in the $g$ and $r$ bands. The mergers are shown by orange crosses and the non-mergers by dark blue plus symbols. The boundary is the dashed black line, with its parameters given in the label, together with the accuracy of the classification using this cut.}
        \label{fig:g-r}
\end{figure}

Moreover, the shape of the histograms in the three central bands $g$, $r$, and $i$ seems to hint at the regions of merging and non-merging galaxies in the \code{skyErr} space. This is very likely what the NN is identifying. Figure \ref{fig:g-r} illustrates not only the clear separation of the classes in the $g$-versus-$r$ plane, but also that simply drawing a boundary line is capable of providing an accuracy of 91.59\%. The line was built by performing a grid search, first for the intercepts using a fixed slope of -1, which is approximately perpendicular to the distribution of galaxies, and followed by a subsequent grid search for the slope with the obtained intercept fixed. A similar boundary was found for $g$ versus $i$ and $r$ versus $i$, with accuracies of 91.16\% and 90.47\%, respectively. Table \ref{tab:boundary} compares the accuracy and the rate of mergers and non-mergers correctly identified using either the NN or the boundary cut. It shows that the boundary is less accurate at identifying the non-mergers than the NN, while it does not lose accuracy for the mergers.

\begin{table}%[h]
    \centering
    \caption{Comparison between the application of the NN or of the boundary cut to the logarithm of the \code{skyErr}. The rows indicate not only the accuracy but also the rate of TPs and TNs for each method when applied on the full training dataset. The NN results correspond to the saved weights of the first cross-validation fold.}
    \begin{tabular}{c c c}
%      Variable & Value \\
    
    \footnotesize Method & \footnotesize NN & \footnotesize Boundary \\
    \hline \hline
      \footnotesize Accuracy &  \footnotesize 92.79 \%  & \footnotesize 91.59 \% \\
      \footnotesize TPs rate & \footnotesize 95.48 \% & \footnotesize 95.04 \%\\
        
      \footnotesize TNs rate &  \footnotesize 90.11 \% & \footnotesize 88.13 \%\\
    \hline \hline
    \end{tabular}
    \label{tab:boundary}
  \end{table}
 
\subsection{Sky error analysis} 
\label{subsect:skyerr_analysis}

All these gathered results and visualizations confirm that the NN favours the input data that explicitly represent the sky error features contrasting between mergers and non-mergers. The best example of this separation is observed in Fig. \ref{fig:g-r}. It indicates that the error in the central bands characterizes the presence of a merging process.

In physical terms, the sky error performance could have a simple explanation. Mergers produce a chaotic flow of material between the components that in some cases cannot be observed unequivocally in the images because it is not bright enough compared to the galaxy itself. This low brightness could be one of the reasons why image recognition, both by humans and by deep learning methods, can fail or can be inconclusive. Nonetheless, these merging traces could still create a detectable signal that only arises in the sky background around the mergers, and it is so low that only the differential analysis or error estimation is able to discern it. However, this has been obtained from our training dataset, which is limited in the type of galaxy mergers, specifically pre-mergers, and deepness of the images.

\subsubsection{Deeper surveys}

Our training dataset covers a specific deepness region defined by the SDSS imaging, and the galaxy's $r$-band magnitude and spectrometric redshift. As we understand it, the sky error method would require the noise around the mergers to be affected by their low-signal regions. Deeper imaging would transform blurred surroundings into sharp boundaries, impairing the method's accuracy. This makes extending the method to deeper data a profound challenge.

In order to estimate the \code{skyErr} method's performance on deeper data, we decided to search for galaxies within our training set that were observed in Stripe 82 of SDSS\footnote{\url{http://cas.sdss.org/stripe82/en/}}. The Stripe 82 area was imaged by multiple scans, providing a magnitude that was twice as deep as the single-pass SDSS frames \citep{2014ApJ...794..120A}. The sky background error available for the Stripe 82 galaxies was calculated using the same pipeline as in SDSS DR7, and therefore DR6. We encountered 208 counterparts through an astrometric match. Out of those sources, we could retrieve the sky error values in counts only for 192 of them, divided into 92 mergers and 100 non-mergers.

We applied the identification methods we have built to this deeper set. The classification of the Stripe 82 galaxies obtained by locating the \code{skyErr} values in the decision boundary provided a 57.81 \% accuracy, and by applying the NN, we obtained a 56.98 $\pm$ 0.35 \% accuracy.

\begin{figure}
    \centering
        \begin{subfigure}{0.48\linewidth}
                \centering
                \includegraphics[width=\linewidth]{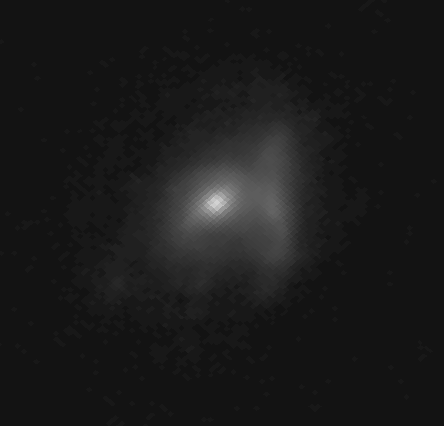}
                \caption{DR6}
                \label{fig:lostmer-DR6}
        \end{subfigure}
        \begin{subfigure}{0.48\linewidth}
                \centering
                \includegraphics[width=0.95\linewidth]{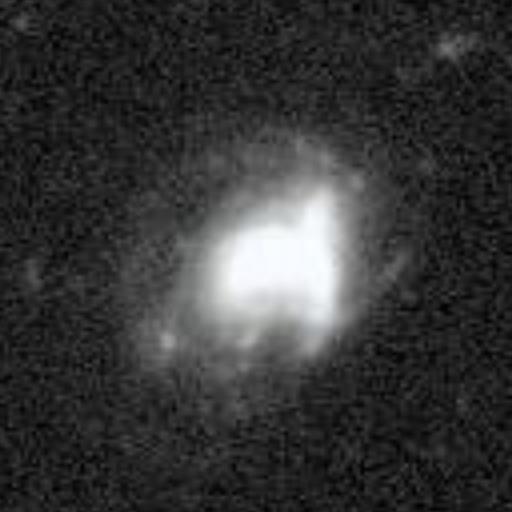}
                \caption{Stripe 82}     \label{fig:lostmer-S82}
        \end{subfigure}  
  \caption{Astronomical frame of a galaxy labelled as a merger from our training dataset. Both images correspond to the $r$-band. On the left side, the DR6 frame is shown, and on the right is the deeper Stripe 82 frame.}
  \label{fig:lostmer_comp}
\end{figure}

We inspected the differences between the DR6 and Stripe 82 merger observations. Figure \ref{fig:lostmer_comp} shows the difference between a merger properly identified in DR6 (Fig. \ref{fig:lostmer-DR6}) but missed in Stripe 82 (Fig. \ref{fig:lostmer-S82}). In this example, the surroundings appear to be more diffuse in deeper data than previously found. This leads us to conclude that the reason why the deepness changes the results is the relative amount of noise and signal in the galaxy's surroundings. Future works will address this issue.

% \subsection{Merger remnants/ post-mergers}%%% I KEEP THIS ORGANIZATION FOR LATER, 
\subsubsection{Merger remnants and post-mergers:}
The \cite{2010MNRAS.401.1043D} catalogue from which we selected the training mergers consists exclusively of merging pairs. As a consequence, we lack post-merging stages in our sample that can indicate whether the sky error method would also identify them or not.

In order to find merger remnants using our current method, we built a catalogue of galaxies in SDSS DR6 within the \textit{r} magnitude and spec-z intervals of our training set. These intervals were [12.24, 18.05] for \textit{r}-mag and [0.01,0.1] for the spec-z, providing up to 286\,616 sources. We then carried out two main studies. First, we located them in the \code{skyErr} decision boundary and visually inspected different regions in the merger's upper half. Second, we made use of the classification in the Galaxy Zoo DECaLS (GZ-D) Campaign 5 \citep{2022MNRAS.509.3966W}. Galaxies with a vote fraction above 0.6 for the classification answer `major disturbance' were defined as post-mergers \citep{2022MNRAS.509.3966W}. Our goal was to visually inspect the astrometric matches with the SDSS DR6 set of these GZ-D post-mergers. We made a lower cut on the number of votes per galaxy to both reduce the inspection time and to make sure they were extensively visualized, avoiding sources that were picked out by their variable retirement rate \citep{10.1093/mnras/stz2816}. The resulting post-merger catalogue contained 45 galaxies.

Among the SDSS DR6 sources we inspected, we did find at least one clear post-merger that had been correctly identified by both the NN and the decision boundary. Among the GZ-D 45 confirmed galaxy post-mergers, only seven of them were found in the merger region. Using the NN, we obtained the same classification. Those galaxies all showed a surrounding material that mixed with the background. Except for the missed merger remnants, the surrounding seemed to be less diffuse than the other seven galaxies.

We discovered that merger remnants can be identified using both the NN and the \code{skyErr} boundary. While not all the GZ-D-based set was recovered, refinement of our method in a future work is likely possible.

\section{Conclusions}

The main goal of this paper was to create an NN and apply it to a class-balanced set of mergers and non-mergers using only photometric information. The dataset is composed of galaxies from SDSS DR6 identified during GZ DR1. The 2$\,$930 mergers from \cite{2010MNRAS.401.1552D} are combined with the same number of non-mergers in GZ DR1 by a nearest-neighbour match in spec-z and $r$-band magnitude. The NN applied is fully connected: it has two layers with 16 neurons whose activation function is ReLU and it has a dropout rate of 0.2; the learning method is Adam with an initial training rate of 5$\times$10$^{-5}$; the output is a softmax probability for a two-class classification; and the classifier is the TensorFlow's BinaryCrossentropy class.

First, the initial research used the band magnitudes, colours, and errors of six different SDSS flux measurement methods. Using the model magnitude, we found a reference accuracy of 68.90 $\pm$ 0.72\%. Checking the other magnitude types brought up the fibre magnitude error importance. The fibre errors reached 83.76 $\pm$ 0.32 \% validation accuracy alone and 88.68 $\pm$ 0.31 \% with bands and colours. Further research showed that the components of the fibre magnitude error could achieve an accuracy of 91.48 $\pm$ 0.31 \%. We found that the parameter that contributes mainly to this high accuracy is the sky background error. We proved that the sky error is able to show differences between mergers and non-mergers that can be identified in the histogram, PCA, t-SNE, and 2D histogram representations, together with the NN results. Finally, we found that the input space of the logarithm of the pre-normalized five-band sky background error in units of counts is able to reach a validation accuracy of 92.64 $\pm$ 0.15 \%. A version of the NN for this last input is published on \code{GitHub}\footnote{\url{https://github.com/LuisEduSuelves/NN16_skyErr-log}} with the saved weights. Moreover, the NN can be substituted by a decision boundary in the planes between the $g$, $r$, and $i$ bands, achieving an accuracy of up to 91.59 \% for the $g$-versus-$r$ plane.

A likely interpretation of this result is that the higher values of the sky background error reflect the traces of merging processes -- for example, faint tidal tails -- otherwise missed by the neural networks due to the dominance of the signal from a galaxy itself. Multi-band analysis of the sky background error additionally makes our network sensitive to the colours of this residual flux that originates from the matter surrounding a merging galaxy.

Future work will concentrate on understanding the sky background error NN and decision boundary applicability zones. It has the potential to identify mergers in multiple stages and in deeper data, but more research in that direction has to be done. We plan on creating catalogues on other SDSS sources, including other SDSS data releases. The next step will be to expand it to other surveys. Another path could be to use deep learning algorithms directly on the sky background maps. Furthermore, we will seek to make a universal implementation of the method for any astronomical research, which would be the final milestone for the method.

\begin{acknowledgement}
We would like to thank the referee for their thorough and thoughtful comments that helped improve the quality of this work. We would like to thank M. Grespan for helpful discussions on this paper. W. J. Pearson has been supported by the Polish National Science Center (NCN) UMO-2020/37/B/ST9/00466 under the project Galaxy Clashes and by the  Foundation for Polish Science (FNP) under the project START. L. E. Suelves and A. Pollo have been supported by the VIMOS Public Extragalactic Redshift Survey (VIPERS) grant from the NCN UMO-2018/30/M/ST9/00757. This research was also supported by the Polish Ministry of Science and Higher Education grant DIR/WK/2018/12. 

Funding for the SDSS and SDSS-II has been provided by the Alfred P. Sloan Foundation, the Participating Institutions, the National Science Foundation, the U.S. Department of Energy, the National Aeronautics and Space Administration, the Japanese Monbukagakusho, the Max Planck Society, and the Higher Education Funding Council for England. The SDSS Web Site is \url{http://www.sdss.org/}.

The SDSS is managed by the Astrophysical Research Consortium for the Participating Institutions. The Participating Institutions are the American Museum of Natural History, Astrophysical Institute Potsdam, University of Basel, University of Cambridge, Case Western Reserve University, University of Chicago, Drexel University, Fermilab, the Institute for Advanced Study, the Japan Participation Group, Johns Hopkins University, the Joint Institute for Nuclear Astrophysics, the Kavli Institute for Particle Astrophysics and Cosmology, the Korean Scientist Group, the Chinese Academy of Sciences (LAMOST), Los Alamos National Laboratory, the Max-Planck-Institute for Astronomy (MPIA), the Max-Planck-Institute for Astrophysics (MPA), New Mexico State University, Ohio State University, University of Pittsburgh, University of Portsmouth, Princeton University, the United States Naval Observatory, and the University of Washington.

\end{acknowledgement}

\bibliography{export-bibtex}
\bibliographystyle{aa}

\begin{appendix}

\section{Reproducing the SDSS fibre magnitudes and errors}
\label{ap:1}

In order to understand what information the fibre magnitude errors enclose, we attempted to reproduce them using the information in the SDSS DR6 documentation\footnote{\url{http://classic.sdss.org/dr6/algorithms/fluxcal.html}}. The documentation indicates that it is calculated as the aperture photometry inside a circle of 3 arc-seconds in diameter, the same angular size as the fibre, after the image is convolved with a 2 arc-second seeing to resemble what the fibre actually sees\footnote{\url{http://classic.sdss.org/dr6/algorithms/photometry.html\#mag\_fiber}}. Therefore, by retrieving the correct catalogues and photometric parameters from the SDSS repository, one could reproduce the fibre errors and examine the properties that make them so relevant for the NN. This appendix is limited to showing how the uncalibrated fibre counts and count errors -- found in the file fpObjc -- relate to the fibre magnitude and errors, together with all the required calibration parameters.

\begin{table*}
        \centering
        \caption{Files employed to recreate the aperture photometry that leads to the fibre magnitude data.}
        
        \begin{tabular}{c c c}
                Name & Type & Contents \\
                \hline \hline
                kfold\_fibre & \small{Input for NN} & \small{Fibre magnitudes and errors} \\
                \hline
                fpObjc & \small{\begin{tabular}{@{}c@{}}Uncalibrated Catalogue, \\ i.e. before converting \\ counts to fluxes\end{tabular}} & \small{\begin{tabular}{@{}c@{}} Fibre raw \code{counts} and \code{counts errors} \\ plus sky background (\code{sky}) and error (\code{skyErr}), \\ interpolated to the source's centre \end{tabular} } \\
                \hline
                drField & \small{Field calibration data} &
                \small{\begin{tabular}{@{}c@{}} Calibration Parameters (per band)
                                \\ (e.g. \code{gain}, \code{airmass}, \code{zeropoints})\end{tabular}}  \\
                \hline
                \hline
                \hline
        \end{tabular}
        \label{tab:files_ref}
\end{table*}

Two main sets of information are required to reproduce the fibre magnitudes: first, the equations that connect the observational measurements with the magnitudes, and second, the files where these measurements are found. Table \ref{tab:files_ref} shows the latter files, and the equations -- as found in the documentation website -- are the following:

\paragraph{Magnitude}
\begin{equation}
        m=-\frac{2.5}{\ln(10)}\left[ \sinh^{-1} \left(\frac{f}{f_02b}\right) + \ln(b)\right] \, \, ,
        \label{eq:mag}
\end{equation}

\paragraph{Magnitude error}
\begin{equation}
        \sigma_m = \frac{2.5}{\ln10}\,\frac{\sigma_{\code{counts}}}{\mathrm{exposure\,time}}\,\frac{1}{f_0}\frac{1}{\sqrt{4b^2 + \left(\frac{f}{f_0}\right)^2}} \, \, ,
        \label{eq:old_E}
\end{equation}

\paragraph{Counts error}
\begin{dmath}
	\sigma_{\code{counts}} = \left[\frac{\code{counts} + (\code{sky}\cdot N_{\mathrm{pixels}})}{\mathrm{gain}} + \\ N_{\mathrm{pixels}}\cdot(\code{dark variance} + \code{skyErr})\right]^{1/2} \, \, .
	\label{eq:old_CE}
\end{dmath}
Here, $f$ is the pixel units in \code{counts} divided by the exposure time; $f_0$ is the zeropoint, that is, the flux of an object with zero magnitude, given the atmospheric conditions and the system's instrumentation; $b$ is the softening parameter that indicates the flux level at which linear behaviour of $\bm{m}$ sets in; \code{sky} is the sky background estimation and \code{skyErr} its error, both interpolated on the source centroid; gain is the telescope's CCD's gain; $N_{\mathrm{pixels}}$ is the size in pixels of the aperture used, and \code{dark variance} is the dark current's variance calibrated for the given frame. It should be noted here that $\code{sky}\cdot N_{\mathrm{pixels}}$ is the sky counts summed over the same area as the object counts, as indicated in the documentation count error\footnote{\url{http://classic.sdss.org/dr6/algorithms/fluxcal.html\#counterr}}

These initial equations did not succeed in reproducing the fibre errors, and some modifications were found to be necessary. In order to justify the modification of these relations, we compared the fibre magnitudes, magnitude errors, and count errors obtained from the original and from our modified formulae. This study was made for ten arbitrary galaxies of the dataset in each of the five pass bands.

 %\newpage

\subsection{Relation of fpObjc fibre counts and count errors with the fibre magnitude and error}

\begin{figure}
        \centering
        \includegraphics[width=\linewidth]{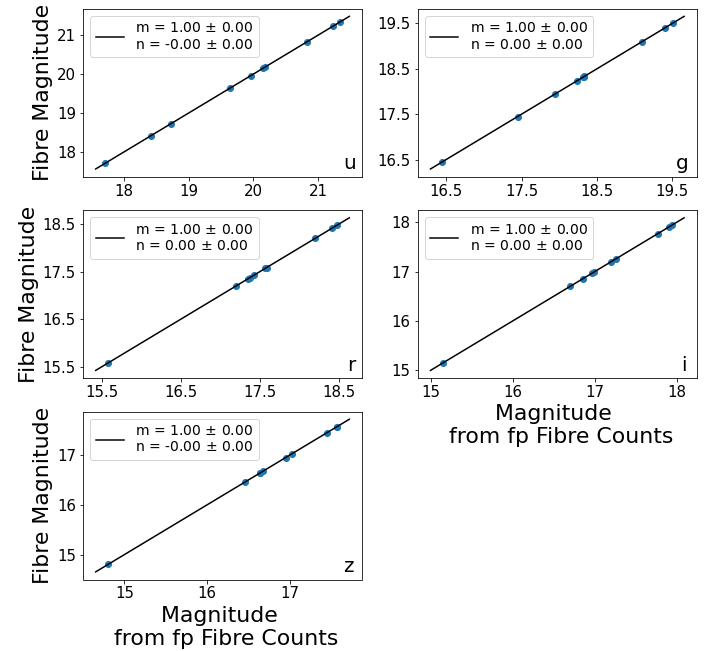}
        \caption{Linear regression between the fibre magnitude extracted from the CasJobs portal (y-axis) compared to the magnitude calculated using Eq. \ref{eq:mag} from the fibre counts (\code{counts}) from the \code{fpObjc} catalogue (x-axis). The fit corresponds to ten galaxies pre-selected from our training dataset, and is done for all five SDSS bands, $u$, $g$, $r$, $i$, and $z$, shown in the five panels.}
        \label{fig:panel-1}
\end{figure}

First, to confirm that the fibre counts (\code{counts}) provided the fibre magnitudes, we took the fibre counts given in the fpObjc catalogue and applied Eq. \ref{eq:mag}. Those fibre counts were the SDSS's counts extracted from the final seeing-convolved frame. Figure \ref{fig:panel-1} gives the five-band panels, where the y-axis shows the fibre magnitude and the x-axis shows the magnitude resulting from the fibre counts. The parameters $f_0$ and $b$ were extracted from the field calibration dataset fpC. The straight black lines indicate the linear fits of the scatter plots, with the fit parameters and their errors in the legend. This linear fit confirms that Eq. \ref{eq:mag} does successfully relate counts and magnitudes.

\begin{figure}
        \centering
        \includegraphics[width=\linewidth]{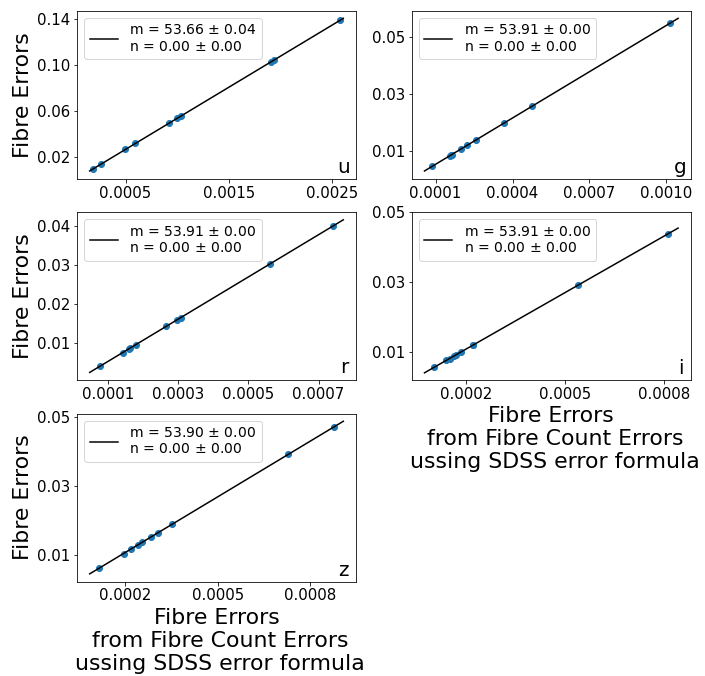}
        \caption{Linear regression, for the same galaxies and bands as in Fig. \ref{fig:panel-1}, between the fibre magnitude errors extracted from the CasJobs portal (y-axis) compared to the fibre magnitude errors calculated using Eq. \ref{eq:old_E} from the fibre count errors from the \code{fpObjc} catalogue (x-axis).}
        \label{fig:panel-3-old}
\end{figure}

\begin{figure}
        \centering
        \includegraphics[width=\linewidth]{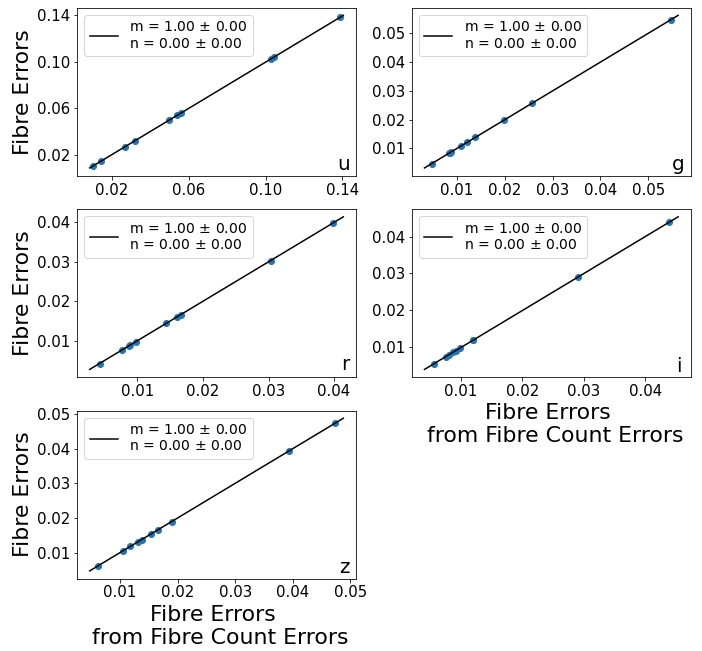}
        \caption{Same as for Fig. \ref{fig:panel-3-old}, but using the new equation for the magnitude errors (Eq. \ref{eq:new_E}).}
        \label{fig:panel-3}
\end{figure}

Second, we applied Eq. \ref{eq:old_E} to the fibre count errors in the fpObjc file. However, the fit showed a slope corresponding to the exposure time of the frames, as shown in the slopes of the five panels in Fig. \ref{fig:panel-3-old}. Therefore, we modified the relation to Eq. \ref{eq:new_E}, and Fig. \ref{fig:panel-3} is the resulting fibre magnitude error. It lacks the exposure time slope and shows a one-to-one relation that confirms the presence of either a typing error or an inconsistent definition.

\paragraph{New magnitude error}
\begin{equation}
        \sigma^{\mathrm{new}}_{m} = \frac{2.5}{\ln10}\,\frac{\sigma_{\code{counts}}}{f_0}\frac{1}{\sqrt{4b^2 + \left(\frac{f}{f_0}\right)^2}} \, \, ,
        \label{eq:new_E}
\end{equation}
in comparison with Eq. \ref{eq:old_E}, the magnitude error is not divided by the exposure time.

\subsection{Count error formula}

\begin{figure}
        \centering
        \includegraphics[width=\linewidth]{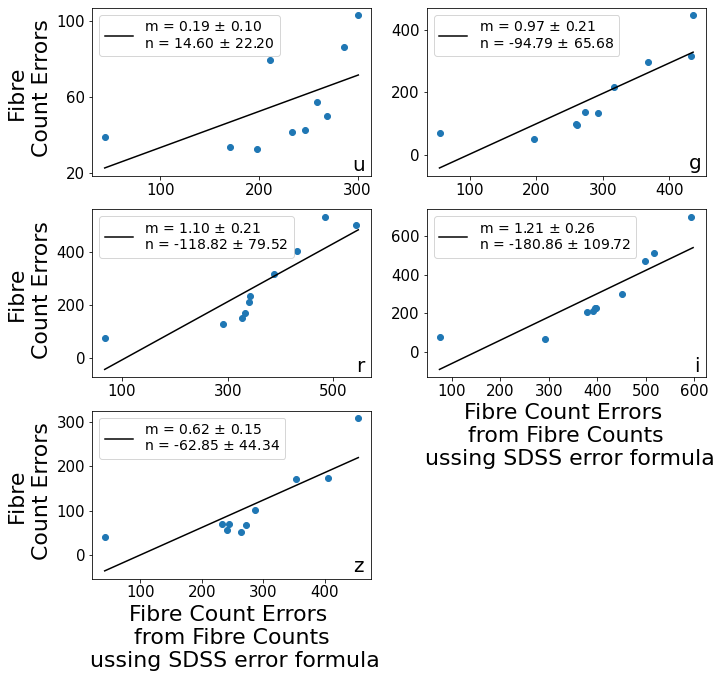}
        \caption{Linear regression between the fibre count errors (y-axis) compared to the fibre count errors calculated using Eq. \ref{eq:old_CE} from the fibre counts (x-axis).}
        \label{fig:panel-5}
\end{figure}

We have confirmed that the counts and count errors in fpObjc lead to the fibre magnitudes and errors, respectively. We have also corrected the exposure time factor in Eq. \ref{eq:old_E}. Nonetheless, when attempting to reproduce the fibre count errors using Eq. \ref{eq:old_CE}, two problems arise. The first one is the ambiguity in the dimensional analysis. On the one hand, the units of \code{skyErr} are counts -- in the fpObjc catalogue, although CasJobs contains them in maggies -- but \code{dark variance} is given in counts squared. The formula is wrongly adding an error with a variance. On the other hand, the first and the second terms differ from each other in the gain denominator. Its units are [gain] = photo-electrons/counts, implying the first term is in counts$^2$/photo-electrons and the second in counts$^2$. This supported applying the correction \code{skyErr}$^2$ over \code{dark variance}$^{1/2}$. The second problem is illustrated in Fig. \ref{fig:panel-5}. The linear fits between the original fibre count errors and those calculated using Eq. \ref{eq:old_CE} are either quite deviated, as appears to be the case in bands $u$ and $z$, or with a big variance and apparently a non-linear shape, as in the $g$, $r$, and $i$ bands.

\begin{figure}
        \centering
        \includegraphics[width=\linewidth]{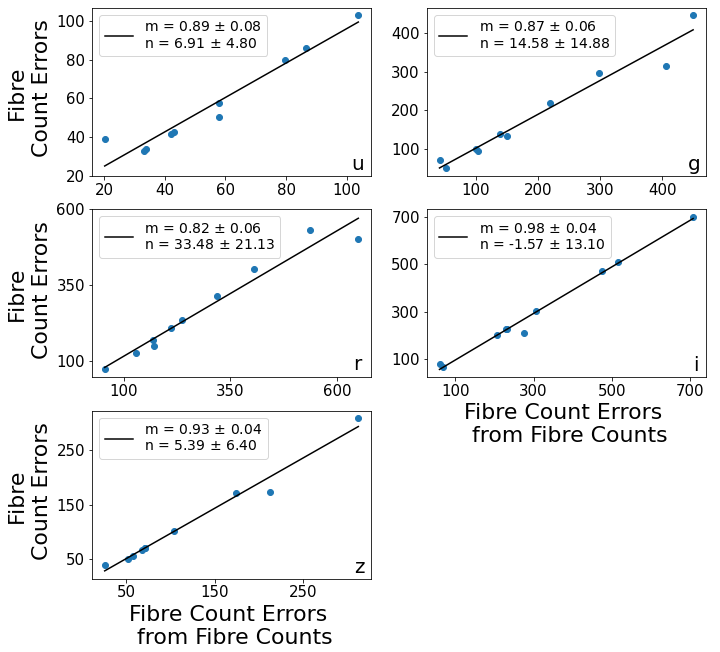}
        \caption{Same as for Fig. \ref{fig:panel-5}, but using the new equation for the count errors (Eq. \ref{eq:new_CE}).}
        \label{fig:panel-2}
\end{figure}

We defined Eq. \ref{eq:new_CE} improving Eq. \ref{eq:old_CE}:
\paragraph{New counts error}
\begin{dmath}
	\sigma^{\mathrm{new}}_{\code{counts}} = \left[\frac{\code{counts} + (\code{sky}\cdot A_{\mathrm{pixels}})}{\mathrm{gain}} + \\ A_{\mathrm{pixels}}\cdot(\code{dark variance} + \code{skyErr}^2) \right]^{1/2} \, \, .
	\label{eq:new_CE}
\end{dmath}

The improvement comes mainly from reducing the dimensional analysis ambiguity in the second term. We also changed the name of $N_{\mathrm{pixels}}$ to $A_{\mathrm{pixels}}$ so that it illustrates better that it is the area covered by the fibre aperture in pixel$^2$ units.

Figure \ref{fig:panel-2} compares the count errors with the result of Eq. \ref{eq:new_CE} on the counts. In contrast to Fig. \ref{fig:panel-5}, a better linearity of the fit can be observed both visually and in the parameter's errors. The slope for all five bands is more uniform, and for the $g$, $r$, and $i$ bands, a value of 1 for the slope is within the error bars, although it still deviates from the identity for $u$ and $z$. Nonetheless, the uniformity of the fits supports Eq. \ref{eq:new_CE}.

\begin{figure}
        \centering
        \includegraphics[width=\linewidth]{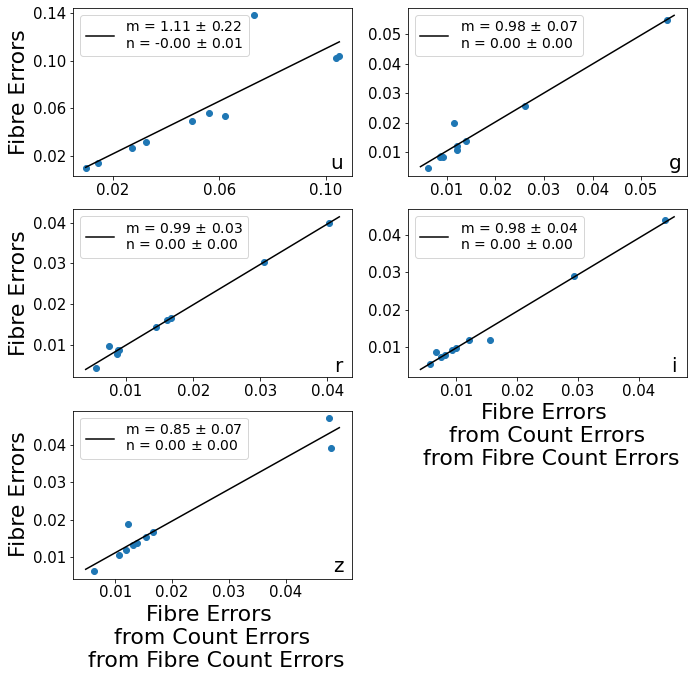}
        \caption{Linear regression between fibre magnitude errors (y-axis), compared to the fibre magnitude errors calculated subsequently using Eqs. \ref{eq:new_E} and \ref{eq:new_CE} on the fibre counts (x-axis).}
        \label{fig:panel-4}
\end{figure}

\begin{figure}
        \centering
        \includegraphics[width=\linewidth]{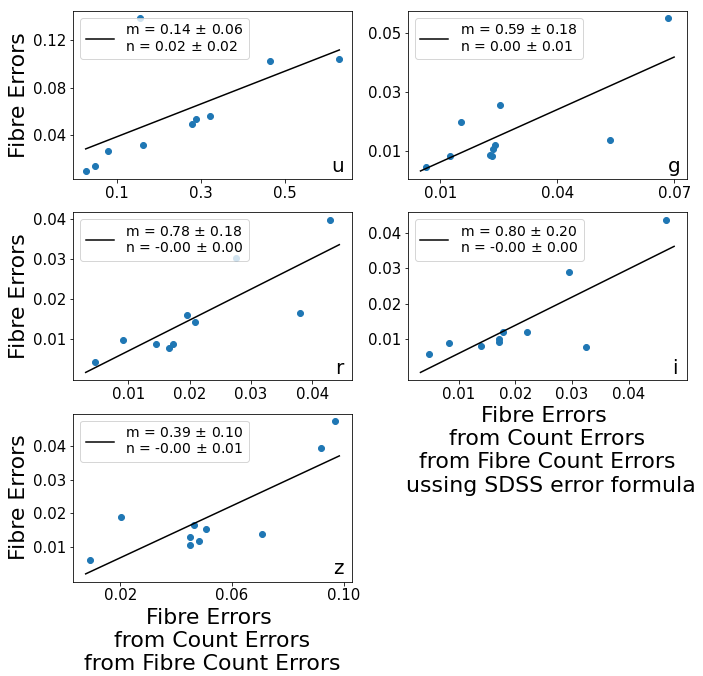}
        \caption{Same as for Fig. \ref{fig:panel-4}, but using the old equation for the count errors (Eq. \ref{eq:old_CE}).}
        \label{fig:panel-6}
\end{figure}

To finalize, using the calculated count errors in the x-axis of Fig. \ref{fig:panel-2}, we applied the new magnitude error formula Eq. \ref{eq:new_E} and compared the result with the fibre magnitude errors in Fig. \ref{fig:panel-4}. The linear fit was quite strong. The intercept was null for all bands and the slope differed from one only for the $z$ band, showing a large relative error only for $u$. Some outliers seemed to spoil the results -- such as the top one in the u-band panel, or the two separated ones in the top right area of the z-band panel. Figure \ref{fig:panel-6} shows, in contrast to Fig. \ref{fig:panel-4}, the magnitude errors when applying Eq. \ref{eq:old_CE}. From the slopes and the visual scatter, it is evident that the fibre errors were incorrect. We note that the intercepts were all almost zero due to the nature of the asinh magnitude formula.

Our purpose in understanding the fibre errors was to identify its inputs to move forwards in Sect. \ref{subsect:fiber_components}. We did not study it further since we considered that the results confirmed that the \code{counts}, \code{sky}, \code{skyErr}, and \code{dark variance} were those inputs.

\end{appendix}

\end{document}